\documentstyle{l-aa}

\input epsf.sty 

\newcommand{\lsim}{\ \raise -2.truept\hbox{\rlap{\hbox{$\sim$}}\raise5.truept\hbox{$<$}\ }}
\newcommand{\gsim}{\ \raise -2.truept\hbox{\rlap{\hbox{$\sim$}}\raise5.truept\hbox{$>$}\ }}

\begin{document}

   \thesaurus{10.15.1; 11.13.1; 11.19.4}

\title{Predicted HST FOC and broad band colours for young and intermediate
Simple Stellar Populations}

   \subtitle{}

\author{
E. Brocato\inst{1,2}
\and 
V. Castellani\inst{3}
\and
G. Raimondo\inst{1,2}
\and
M. Romaniello\inst{4,5,6}
                   }

   \offprints{E. Brocato}

\institute{
Osservatorio Astronomico di Collurania, Via M. Maggini, I-64100 Teramo, 
Italy\\
email: graimondo,brocato@astrte.te.astro.it
\and
Istituto Nazionale di Fisica Nucleare, LNGS, I-67100 L'Aquila, Italy
\and
Dipartimento di Fisica dell'Universit\`a di Pisa, Piazza Torricelli 2, I-56126 Pisa, Italy\\
email: vittorio@astrte.te.astro.it
\and
Scuola Normale Superiore, Piazza dei Cavalieri 7, I-56126 Pisa, Italy 
\and
Space Telescope Science Institute, 3700 San Martin Drive, Baltimore, MD 21218 
\and 
European Southern Observatory, K. Schwarzschild Str 2, Garching b. 
M\"unchen, D--85748, Germany\\
email: mromanie@eso.org 
}

\date{Received ...; accepted ...}

   \maketitle
\markboth{ }{}
   
\begin{abstract}

This paper  presents theoretical HST and broad band colours 
from  population synthesis models  based on an homogeneous set 
of stellar evolutionary tracks as
computed under canonical (no overshooting) assumptions, covering
the range of cluster ages  from $t~=~8$ Myr to $t~=~5$ Gyr for three 
different metallicities (Z$~=~0.02$, $0.006$, and $0.001$). 
Statistical fluctuations in the cluster population have been
investigated, assessing the predicted fluctuations of the
various colours as a function of the cluster integrated absolute
magnitude.
We show that the red leak in HST UV filters  deeply affects the 
predicted fluxes and colours. However, we find that for 
F152M$-$F307M$\le 0.5$ and for F170M$-$F278M$\le 0.5 $ (which 
means ages lower than 1 Gyr) the HST UV colours can still be used 
to infer reliable indications on the age of distant clusters. 
Moreover, one finds that 
the age calibration of these colours is scarcely affected by 
the amount of original helium or by the  assumed IMF. 
On this basis, we present a calibration of the 
HST UV two-colours (F152M$-$F307M~vs~F170M$-$F278M) in terms of cluster
ages for the three above quoted metallicities.
We suggest the combined use of HST UV colours and IR colours (V$-$K in 
particular) to disentangle the metallicity-age effect in 
integrated colours of young stellar populations ($t \le 1 Gyr$).


\end{abstract}

\section{Introduction}

The integrated light emitted by distant stellar populations is among
the most stimulating evidence concerning the evolution of matter in
the Universe. The challenge of interpreting integrated spectra and/or
colours of distant objects (galaxies and stellar clusters) in terms of
stellar ages and chemical compositions was early gathered by several
authors, approaching the problem of population synthesis both from a
theoretical or an observational point of view (see, e.g., Larson 1974;
Tinsley 1980; Searle, Wilkinson and Bagnuolo 1980; Renzini and Buzzoni
1986; Barbaro and Olivi 1986; Arimoto and Yoshii 1986, 1987;
Rocca-Volmerange and Guiderdoni 1987).  According to the relevance of
the subject, population synthesis models have been continuously
upgraded over the years (Worthey 1994; Bressan et al. 1994; Leitherer
\& Heckman 1995; Bruzual \& Charlot GISSEL96 as quoted in Leitherer et
al. 1996; Vazdekis et al. 1996) and a wide data base collecting the
most recent results can be found in Leitherer et al. (1996).
This scenario has been recently extended to HST photometric bands
by Padua group (Chiosi et al. 1997).

As pointed out by Charlot et al. (1996), the adopted library of stellar
evolutionary tracks is the basic assumption which determines
the predicted integrated properties of a Simple Stellar Population,
i.e., of a group of coeval and chemically homogeneous stellar objects 
(hereinafter SSP). In a previous work (Barbero et al. 1990) we collected 
evolutionary tracks from several papers in the literature to present
an age calibration of integrated UV colours of young stellar clusters in the
Large Magellanic Clouds. In that paper we found a rather satisfactory
agreement between theoretical predictions and observation. On this basis 
we suggested the use of the UV two-colours diagram (C(15$-$31) vs C(18$-$28)) 
as a useful test for theories, to be calibrated in terms of cluster ages.   
After that paper, the Teramo-Pisa-Frascati (TPF) cooperation  has progressively
made available  a homogeneous set of stellar evolutionary tracks which 
covers the main evolutionary phases expected in the stellar populations
over a rather large range of assumed ages and chemical compositions. 
In this paper  we will make use of this opportunity to revisit the problem
of cluster integrated light by relying on such a self-consistent and
improved theoretical scenario.

As already discussed in several papers (Brocato et al. 1994 and
reference therein) TPF isochrones differ from  similar computations 
based on Padua or Geneva library of stellar evolutionary models 
(see Leitherer et al. 1996 for the precise references of these tracks) 
mainly because in TPF stellar models the efficiency of 
overshooting from convective hydrogen burning cores is assumed negligible while
both Padua and Geneva models adopt a relatively high efficiency, 
directly leading to different evolutionary times of the 
core H-burning phase. Since the actual amount of core overshooting 
is still far from being a settled question and since most of the 
population synthesis models  are based on tracks  
computed by using Padua or Geneva tracks, it is obviously interesting
to discuss the results of population synthesis as computed in the limit
of no overshooting, at least to allow a comparison between alternative
evaluations on that matter. 
In this context one has to notice that 
all the  referred evolutionary computations rely on similar input 
physics, allowing a meaningful comparison among different results.

The main goal of this paper is to address some new approaches concerning the
use of population synthesis results. As a first point, 
we will discuss in some details not only the
uncertainties connected to the assumed Initial Mass Function (IMF) 
of cluster stars, but also the statistical fluctuations connected 
to the richness of the cluster in terms of abundance of 
luminous stars. We will show that these fluctuations can be easily put
in dependence of the total integrated cluster light, thus
providing a useful recipe to estimate the degree of
reliability of observational data from SSP.
As a further point, the {\it Hubble Space Telescope} (HST) is presently a
fundamental  source of high quality data for distant galaxies and stellar
clusters. 
Since HST works in space, it can access the UV wavelength range, which 
has been already recognized 
as a fairly good indicators of the age of stellar populations 
(Barbero et al. 1990; Cassatella et al. 1996; Chiosi et al. 1997),
at least in the range  from $\sim$ 10 Myr to a few Gyr.
According to such evidence, we will present predicted integrated colours based 
on HST filters as expected by simple stellar populations with
$2\cdot 10^{-2} \ge Z \ge 1\cdot 10^{-3}$ and ages 
ranging from $ t~=~8\cdot 10^6$ yr up to $ t~=~5 \cdot 10^9$ yr,
the upper limit of ages being chosen in order to keep our simulations 
free from assumptions about the temperature distribution of 
Horizontal Branch (HB) stars. Thus theoretical predictions are
primarily intended to investigate the integrated colours of not-too-old
stellar clusters.

The next section is devoted to present the theoretical background and 
to discuss  the choice of the selected filters. Integrated colours for a 
reference population are discussed in the following section. The influence of 
IMF and chemical composition on the synthetic integrated colours
 is then discussed in section 4.
The comparison with previous models and with LMC clusters is given in section
5.  The age calibration and the features of a UV colour-colour diagram are 
presented in section 6, where we approach the problem of deriving age
{\it and}  metallicity from integrated colours together with a tentative
application of present results. Final remarks and conclusions close the paper.

\section{The theoretical background}

A major requirement for population synthesis models concerns 
the set of stellar evolutionary tracks adopted to predict the 
effective temperature and the luminosity of stars 
contributing to the total energy flux. Here we will rely on the
extensive set of stellar models presented for high (Brocato \& 
Castellani 1993), intermediate (Cassisi, Castellani \& Straniero 1994) 
and low mass (Straniero \& Chieffi 1991; Castellani, Chieffi \& 
Straniero 1992) stars. This set  provides an homogeneous and 
complete evolutionary scenario since all the computations 
have been performed with the same stellar evolutionary code and 
with similar physical assumptions. It covers a wide range of 
values of both stellar masses and/or chemical compositions. 
As an example, Fig. 1 shows the run of evolutionary tracks for 
Y$~=~$0.27 and Z$~=~$0.02. 

\begin{figure}[htb]
\epsfxsize=8.8cm 
\hspace{3.5cm}\epsfbox{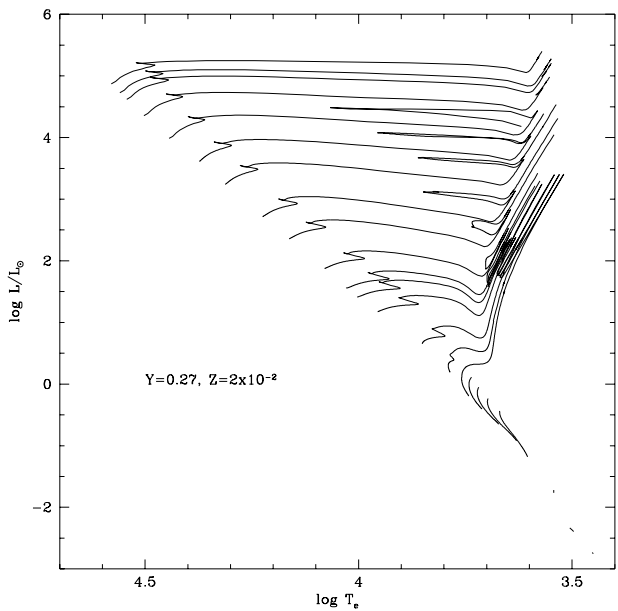} 
\caption[h]{The adopted set of stellar evolutionary tracks
for Y$~=~$0.27 and Z$~=~2 \cdot 10^{-2}$.
The plotted masses are 0.6, 0.7, 0.8, 0.9, 1.0,
1.2, 1.5, 2.0, 2.3, 2.5, 3.0, 4.0, 5.0, 7.0, 9.0, 12, 16, 20, 22, 25 $M_{\odot}.$}
\label{fig1}
\end{figure}

The adopted grid of stellar evolutionary tracks has been 
already submitted to  extensive comparisons with observations, 
which have shown a reasonable accuracy of the models in reproducing 
the behavior of real stars in terms of lifetimes, effective temperatures 
and luminosities for a large range of ages. 
Colour-Magnitude (CM) diagrams of stellar clusters as NGC 2004
($t~=~8\cdot 10^6$ yr: Bencivenni et al. 1991), NGC 1866 and NGC 1850
($t\simeq 1\cdot 10^8$ yr: Brocato et al. 1989 and Gilmozzi et al. 1994),
 M11 ($t~=~1.5\cdot 10^8$ yr: Brocato, Castellani and Di Giorgio 1993), 
other open galactic clusters (Castellani, Chieffi and Straniero 1992)
and  NGC 188 (Caputo et al. 1990) all show to be in substantial agreement
with the quoted theoretical models concerning magnitudes, colours and
stellar counts in the various evolutionary phases.

To compute integrated colours we updated the population 
synthesis code presented by Brocato et al. (1990). The major changes are
related to the new set of evolutionary tracks  and 
to the relation adopted to derive magnitudes and colours from the theoretical 
data log(L/L$_{\odot}$), log(T$_{e}$).
The input parameters are the age (t), helium content (Y), 
metallicity (Z) and total number of stars (N) and slope $\alpha$ of the
Initial Mass Function (IMF), assumed to be a power law. A value of
$\alpha=2.35$ corresponds to the classical Salpeter (1955) slope
The original chemical composition (Y and Z) defines the set of tracks 
which is then used to compute the synthetic CM diagram.
In all the models, we use 
a Monte Carlo simulation to generate the mass of each 'star' 
according to the power law distribution, simulating in this way the stochastic 
behavior of the IMF.  Finally, a simple sum of the flux of 
each star provides the integrated fluxes and colours. 
This procedure has the advantage of keeping under control the 
results of the simulations by showing the 'theoretical' CM distribution 
of the simulated population that can be compared to the 
observed CM diagram of real stellar clusters for a further 
check. Moreover, the contribution of each evolutionary phase can be
easily evaluated either in terms of contribution to the integrated 
bolometric magnitude or to the integrated magnitude in a given filter.  
\par

The choice of the filters has been based on the three following considerations:

1. the major age indicator in CM diagrams of young and intermediate 
age populations is the upper and {\it hottest} Termination of the 
Main Sequence (TMS). The stars located at the TMS are the most 
significant contributors to the integrated light for those populations. 
This means that the most efficient region of the spectral energy 
distribution to investigate the age is the UV side. Moreover, a 
not negligible aspect is that the MS phase is also the most 
populated evolutionary phase (thanks to the large  H-burning 
timescales), making this indicator the best choice also for 
statistical reasons.
Barbero et al. 1990 presented a two UV colour diagram (C(15$-$31) vs 
C(18$-$25)) based on the photometric bands explored by the ANS-satellite, 
showing a fair correlation with the age. More recently, 
Cassatella et al. (1996) proved that ages obtained with these integrated 
UV colours are consistent with ages derived by isochrone fittings for 
a sample of LMC clusters. A survey on the available 
HST filters discloses that F152M, F170M, F253M, F278M and F307M
may represent a valid counterpart for the quoted  "ANS" filters. 

2. The V filter is one of the most common filters and it has been 
extensively used also by HST observers. Moreover Dorman et al. 
(1993, 1995) have already shown that it could be very useful in investigating
chemical composition when used in conjunction with UV filters.

3. Standard U, B, R, I, J, K, L and HST-WFPC2 F606W filters have been also
selected because of their extended use with the Wide Field Planetary Camera~2
and NICMOS on board HST as well as in ground-based 
observations of stellar systems. It is thus possible to compare 
population synthesis models with wide band observations of 
distant stellar objects. 
Integrating in such a way the evaluation already given by Chiosi et al. (1997)
conserning the HST expectations.

According to these choices, our population synthesis code 
evaluate colours on the basis of stellar atmosphere model by 
Kurucz (1979ab : K79). This choice relies on the evidence that 
the mixing length parameter adopted in the computations connected to K79
gives a better (but not perfect) approximation of the observed colours of
cool stars than new model atmospheres do (see the discussion in 
Brocato, Castellani \& Piersimoni 1997). Adopting more recent model 
atmospheres (for example as given by Kurucz 1992) would only increase the
discrepancy between predicted and observed magnitudes and colours.

In order to translate the theoretical isochrones from the HR diagram to the
different colour--magnitude diagrams (CMDs), and to calculate the 
integrated colours by simply summing the contribution of all the 
stars, we used the standard HST synphot task running under 
the IRAF package. We computed the (M$_i$--V) colours, where the 
index $i$ stands for the various filters mentioned above, as expected 
from the Kurucz model atmospheres for a wide range of temperatures 
(3750~K $\leq $~~T$_{e}~~\leq$50000~K) and gravities
(0.75~$\leq$~log g~$\leq$~5 in cgs units), covering the excursion of these
quantities during the whole stellar life. 

The adopted model grid has a lower mass limit of M$=$0.6 M$_{\odot}$. However,
one already knows that lower masses
should  give a negligible contribution to the cluster light. 
As a test, we performed a set of numerical experiments 
implementing  the grid of Reference Frame (RF) models (see next section for
definition) with the stellar models of very low mass stars by Alexander et al. 
1996, extending  the lower mass limit down to 0.15 M$_{\odot}$.
The resulting integrated colours are, for all the ages, within 
the expected statistical fluctuations (see below) of the RF models.
As expected, the larger difference is found for the V$-$K colour 
($\Delta (V$-$K) \le 0.1$ mag), due to the very low temperature 
of the faint but numerous lower main sequence stars. 
However note that the lower portion of the MS should play a not negligible role
when dealing with colours at even larger wavelengths, i.e., 
in the far IR. 	

\section{ Integrated colours for the Reference Population}

To present and to discuss the results of theoretical
simulations we will assume as a reference frame (RF) the results
concerning a stellar population with
solar composition (Y$~=~$0.27, Z$~=~$0.02) where a total number N$~=~$30000 of
stars is distributed according to a Salpeter IMF ($\alpha~=~$2.35)
between 0.6 and 25 M$_{\odot}$. After discussing theoretical 
predictions concerning such a sample, we will refer to this 'archetype' 
to investigate the influence of changing the assumptions 
either on the IMF or on the chemical composition.

\begin{figure}[tb]
\epsfxsize=8.8cm 
\hspace{3.5cm}\epsfbox{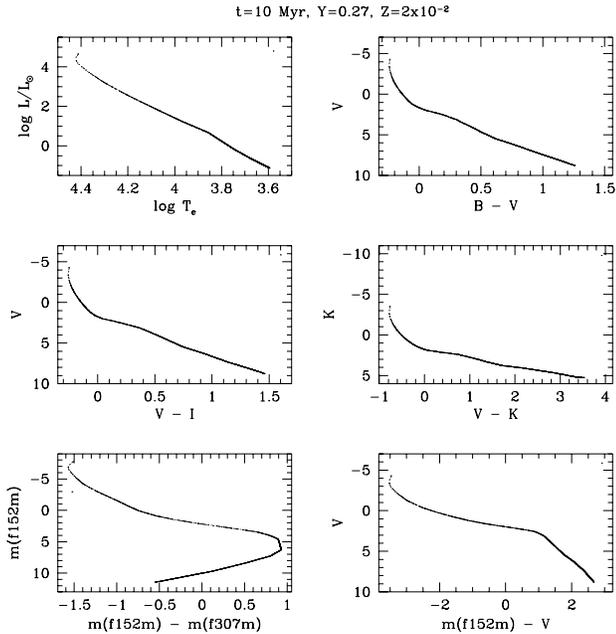} 
\caption[h]{Theoretical log(L/L$_{\odot}$), log(T$_{e}$) RF model for an age of 
$t~=~10$ Myr  (upper left panel) compared with the
corresponding predicted  CMD for selected filters.}
\label{fig2}
\end{figure}

\begin{figure}[htb]
\epsfxsize=8.8cm 
\hspace{3.5cm}\epsfbox{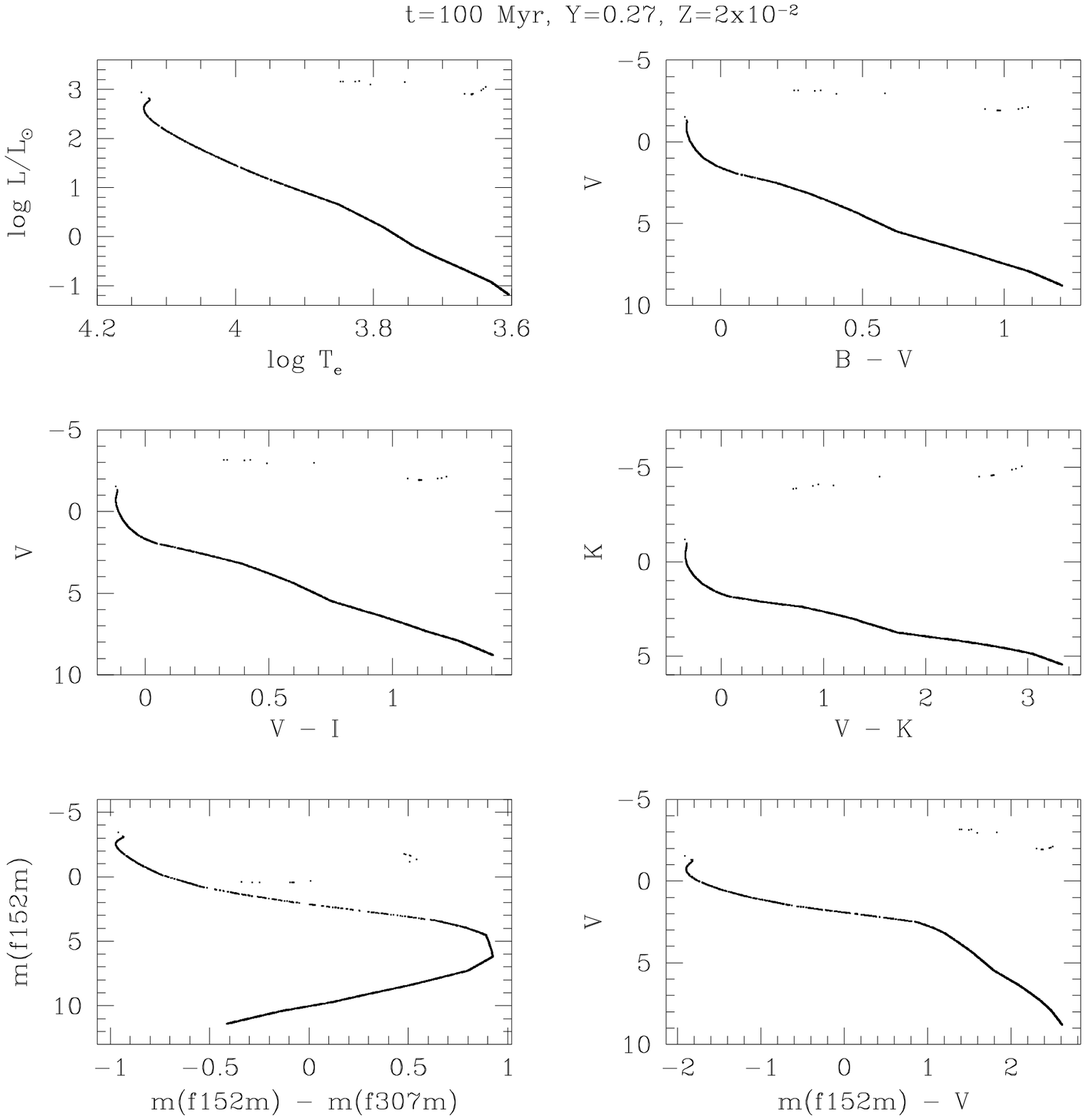} 
\caption[h]{As Fig. 2 but for $t~=~100$ Myr.}
\label{fig3}
\end{figure}

\begin{figure}[tb]
\epsfxsize=8.8cm 
\hspace{3.5cm}\epsfbox{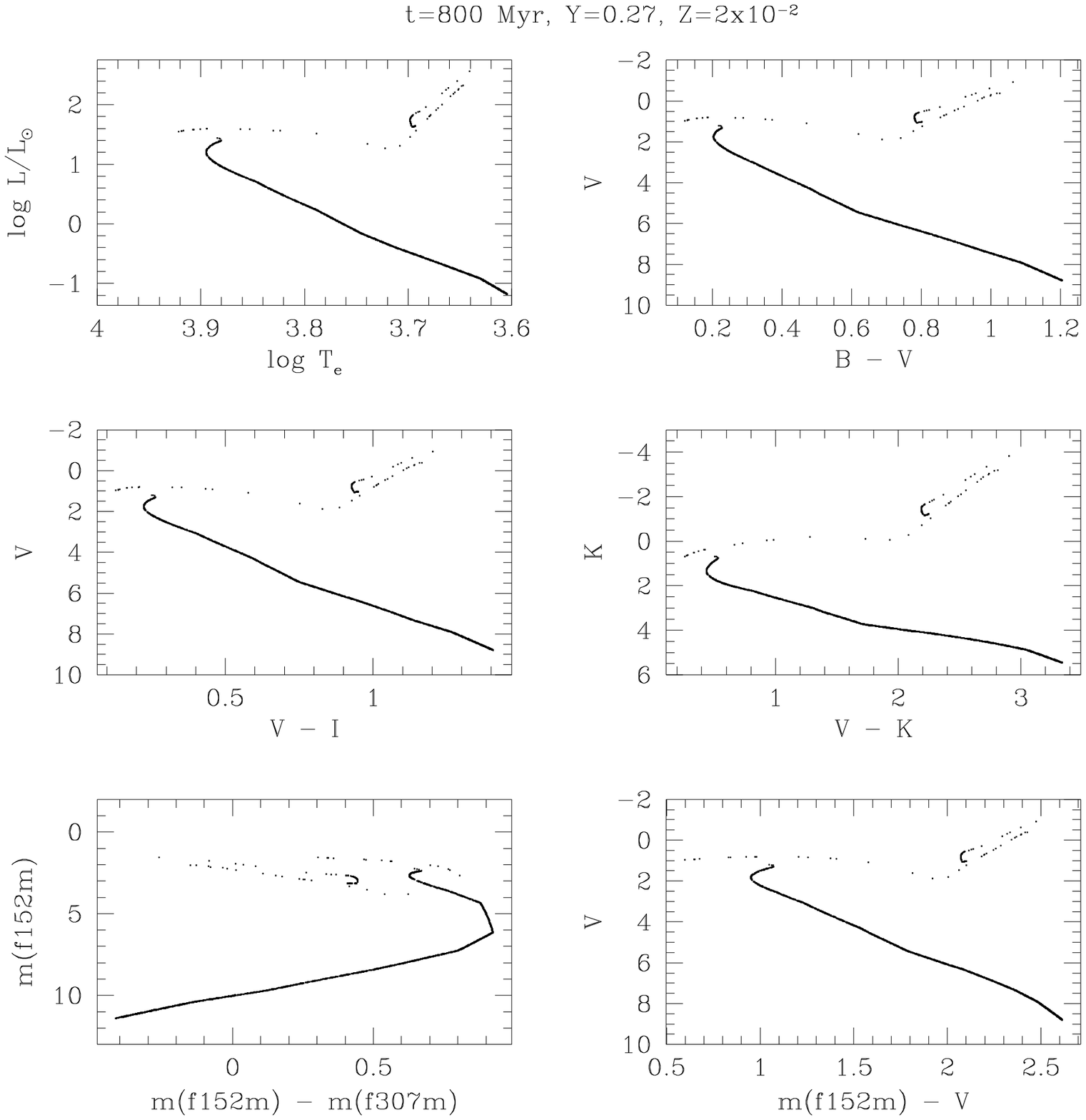} 
\caption[h]{As Fig. 2 but for $t~=~800$ Myr.}
\label{fig4}
\end{figure}

\begin{figure}[tb]
\epsfxsize=8.8cm 
\hspace{3.5cm}\epsfbox{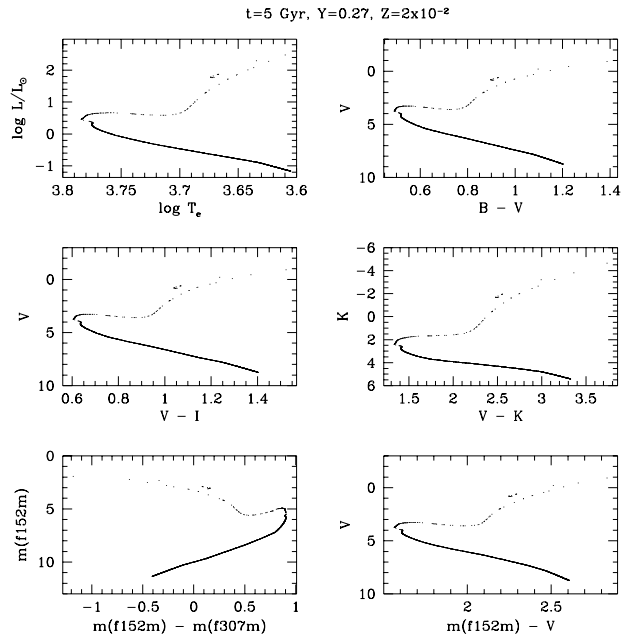} 
\caption[h]{As Fig. 2 but for $t~=~5$ Gyr.}
\label{fig5}
\end{figure}

To begin with, we present a selected sample (top-left panels of Fig. 2 - 5) of 
the theoretical logL/L$_{\odot}$ vs log T$_{\rm eff}$ diagrams
as obtained adopting (for graphic reasons) N$~=~$3000 and for
four representative ages ($10^7,\ 10^8,  \ 8\cdot 
10^8,\ 5\cdot 10^9$ years). 
Bearing in mind that the 'stars' plotted into  theoretical diagrams represent 
the contributor to the total integrated  flux of the population, 
the figure shows the well known occurrence
for which young clusters are dominated by hot giants, whereas
for larger ages the flux from Red Giant and Asymptotic Giant Branch
stars begins dominating. Theoretical expectations
about broad band colours can be better understood
by looking at Fig. 2 to Fig. 5 where we compare, for each selected
cluster age, the theoretical logL/L$_{\odot}$ vs log T$_{\rm eff}$ diagrams
with similar diagrams but for  selected photometric bands.

One should in particular notice the curious CMD of stellar populations
of different ages, as seen by HST red leaked filters (see also Chiosi et al. 1997). 
One finds
that in the  UV CMD (F152M~vs~F152M$-$F307M) the MS discloses 
an unusual turn back at F152M$-$F307M$\simeq 0.9$ which 
means that, in such filter system, faint cool MS stars have a 
colour very similar to the stars populating the upper portion of the MS.
Note also that  cool core He-burning stars have rather 'blue' colours.
Both these effects will strongly influence the expectations about
cluster integrated light. 

Bearing in mind such a scenario, we present in Fig. 6 (Table 1a,b)
theoretical expectation about cluster integrated colours as obtained
from cluster populated by 30000 stars between 0.6 and 25 M$_{\odot}$.
The labeled errors show the 1 $\sigma$ dispersion of the results
obtained in 100 independent simulations. One can note that all the HST 
UV colours disclose monotonic relationships with the age, up to $10^9$ yr, 
thanks to the fact that most of the flux emitted at these wavelengths is 
generated by the more luminous main sequence stars.

\begin{figure}[htb]
\epsfxsize=8.8cm 
\hspace{3.5cm}\epsfbox{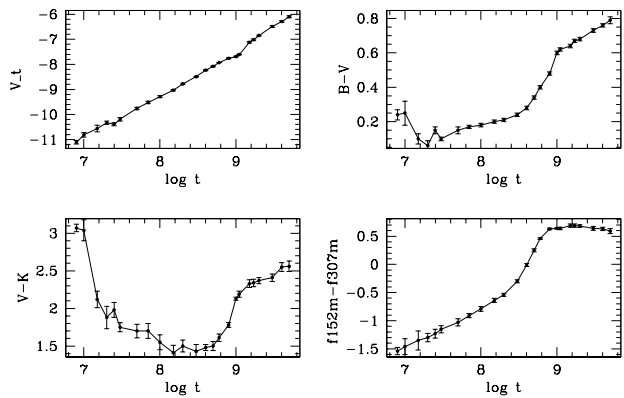} 
\vspace{-3cm}
\caption[h]{Time evolution of integrated V magnitude and
selected colours of the RF models (see text).}
\label{fig6}
\end{figure}

At $\log t \simeq 9$ the HST UV colours loose their sensitivity to 
variation in age and become roughly constant. This is not
due to the variation in the UV flux of the population, 
but it is the result of the red leak of the HST filters which 
transmit the flux emitted by RGB stars and by the numerous low MS stars. 
This can be seen in Fig. 7, where the colour expected by HST filters is
plotted against the colour obtained by theoretical filters centered at 
similar wavelengths and with a passband of 200~\AA, but without red leak.
The red leak, then, acts in the way of simulating the presence of 'blue'
stars in the UV CMD. However, the hot stars of the MS termination are
brighter  than the 'redleaked' cool stars 
up to  $\log t \simeq 9$. For this reason the relation HST UV integrated 
colours vs age shown in Fig. 2 holds up to this age and becomes almost 
flat for larger ages. In conclusion, the previous discussion indicates 
that the red leak plays a relevant role when interpreting  HST UV 
integrated colours.

\begin{figure}[htb]
\epsfxsize=8.8cm 
\hspace{3.5cm}\epsfbox{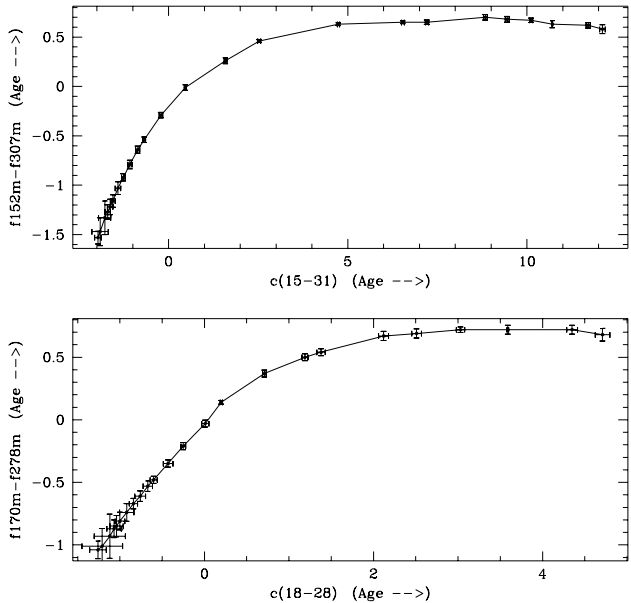} 
\caption[h]{Correlation between  HST-UV  
and  Barbero et al. (1990) integrated colours. The arrows in the labels
indicate the direction running from low to large ages. Note the red leak 
effect in HST-UV colours.}
\label{fig7}
\end{figure}

Coming back to Fig. 6, one finds that all the broad band colours show 
a relatively flat minimum, i.e. a bluer  colour, at intermediate ages. This 
is due to the  occurrence in the younger population of red supergiants 
experiencing their 
He-burning phase. In particular, very young populations ($\simeq 10^7$ yr)
are expected to have even redder V$-$K values than very old one ($ \ge 10^9$) 
yr. This result will be further examined 
in discussing the effect of metallicity on present models. 
Another interesting feature of the broad band colours is the change of slope 
decreasing the age at $\log t \simeq  8.6$ due to the appearance of 
the Red Giant Branch which leads to redder  colours. 

Before closing this section, we notice that the total abundance of stars
can play a significant role in determining the integrated colour of 
a stellar cluster. Poorly populated clusters
should be affected by large statistical fluctuations in the 
distribution of luminous stars in the CM diagram, which in that case is no 
longer led by the evolutionary constraints, but  governed by 
stochastic rules. We have already found that N$~=~$30000 gives
satisfactorily small fluctuations. However, to have more light 
on such an occurrence, we explored the behavior 
of the RF population at 10 Myr, 100 Myr and 1 Gyr and for different 
total numbers of stars (N$~=~$100, 500, 1000, 3000, 7500, 10000, 30000, 
45000) by computing a series of 100 models for each given age and N value. 
The top left panel in Fig. 8 shows, for each N value, the
1 $\sigma$ dispersion of the expected cluster integrated V magnitude as
computed for the three selected cluster ages. The other panels in the
same figure show theoretical expectations about cluster
integrated colour {\it given as a function of the integrated V magnitude of
the cluster} through the relation depicted by the top left panel. 
As expected, for each given V magnitude, one finds 
that decreasing the 
cluster age integrated colours appear more and more affected by 
statistical fluctuations, as a consequence of the stochastic contribution
from few giants stars in a rapid evolutionary phase.

\begin{figure}[htb]
\epsfxsize=8.8cm 
\hspace{3.5cm}\epsfbox{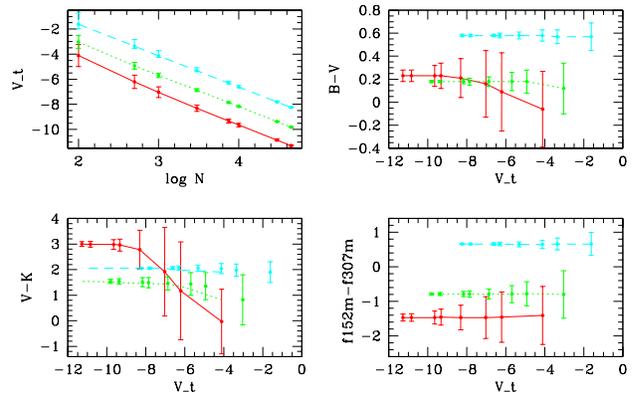} 
\vspace{-3cm}
\caption[h]{The integrated V magnitude 
as function of the total number of stars  populating the RF model
(top left panel) and a selected sample of integrated colours  
as a function of the cluster integrated V magnitude (see text).
Each panel shows the result of simulation for three selected
ages $t~=~10$ Myr ({\it solid line}), 
$t~=~100$ Myr ({\it dotted line}) and $t~=~1$ Gyr ({\it dashed line}).}
\label{fig8}
\end{figure}

As an use result, one finds -- e.g. -- that for an age of 10 Myr the 
Johnson colours B$-$V,  V$-$K (and V$-$I) 
do not correlate with cluster evolutionary
status unless the cluster is brighter than about V$~=~-$9. This is
not the case for UV HST colour, which are much less affected by the
stochastic occurrence of red giant stars. As a whole, data in Fig. 8
give  a useful warning on the use of integrated
colour of stellar clusters.
Bearing in mind these results, in the following  we will limit our 
study to populations for which the MS is 'dominated' by the IMF 
law, discussing in all case the result obtained
from cluster populated by 30000 stars between 0.6 and 25 M$_{\odot}$.

\section{IMF and chemical composition}

To study the influence of the initial distribution of masses,
we have repeated the simulations described above, this time varying the IMF
power law exponent $\alpha$  over a wide range of values. Fig. 9 
shows synthetic colours for $\alpha~=~0$, corresponding to the extreme limit 
of a uniform mass distribution, $\alpha~=~1.35$, i.e. less than 
Salpeter's value, and $\alpha~=~3.35$,  corresponding to a  
distribution steeper than Salpeter's one. At first glance, 
one would expect that decreasing  the
exponent the cluster becomes brighter, since a larger fraction of
stars is pushed toward the more massive and, thus, more  luminous stars. 
The upper left panel in Fig. 9 shows that is not always the case: for ages
larger than, about, 1 Gyr the extreme case of a flat IMF results in fainter
clusters, just because for such large ages the luminosity is produced by the
now depopulated range of less massive stars.

\begin{figure}[htb]
\epsfxsize=8.8cm 
\hspace{3.5cm}\epsfbox{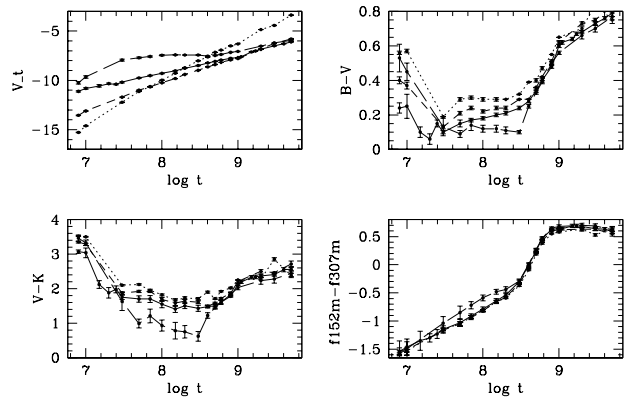} 
\vspace{-3cm}
\caption[h]{The integrated V magnitude and colours 
for different assumption of the IMF exponent: solid line is $\alpha~=~2.35$,
dotted line is $\alpha~=~0.0$, short dashed line is $\alpha~=~1.35$,
long dashed line is $\alpha~=~3.35$.}
\label{fig9}
\end{figure}

The general trends of integrated colours do not appear dramatically  
affected by IMF: the HST UV colours maintain their monotonic behavior 
and broad band colours still show a minimum around $\log t~=~8$. 
However, a large variation in the IMF slope can influence the absolute 
values of the integrated colours depending on the age and on the 
particular colour examined. In general, one finds that the variations 
are smaller than 0.2 mag and 
only in the case of $\alpha~=~3.35$ the colours show larger fluctuations.
However, for $\log t \ge 8.5$ 
the colour vs age relations do not depend on the IMF slope any more, because 
the colours are dominated by RGB and post RGB stars, whose distribution
is fixed by evolutionary timescales only. For younger ages the 
influence of the IMF on the integrated colour increases since 
a non negligible portion  of the emitted flux comes from  MS stars, 
according to their relatively long evolutionary timescales and, thus,
according to their IMF distribution. As a relevant point, one finds that HST
UV colours appear  affected only in the extreme case of a very steep IMF
($\alpha~=~3.35$), showing to be -- in this respect -- a rather robust
indicator of cluster age.

The influence of chemical composition  has been investigated by 
computing selected models either keeping fixed the Helium 
content (Y$~=~$0.27) while varying Z to selected values (Z$~=~2\cdot 10^{-2}, 
6\cdot 10^{-3}, \ 
10^{-3}$) or changing Y (Y$~=~0.23$ and Y$~=~0.27$) for a fixed Z (Z$~=~10^{-3}$).
As far as the metallicity is concerned (Fig. 10, Table 1--3), 
one finds the interesting feature
for which decreasing the metallicity the relation between broad band colours 
and age becomes more and more monotonic. 
For young ages, the Z$~=~$0.001 models show much bluer colour (even 3 mag 
of difference in V$-$K colour)
than the solar metallicity models. This is due to the difference 
in the evolution of intermediate mass stars in the phase following 
the exhaustion of H in the center. Solar metallicity models run 
to the red portion of HR diagram and set their He-burning phase 
at low temperature, near the Hayashi track, for at least half of their 
evolutionary time. On the other hand low metallicity stars (Z$\le$ 0.001)
do not reach this part of the HR diagram, but stay in the blue side 
during all the He-burning phase. We note that such a behavior
is strongly dependent on the treatment of convection in the more
massive models, so that colours of young population appear sensitively
model dependent (see Brocato \& Castellani 1993). 

\begin{figure}[htb]
\epsfxsize=8.8cm 
\hspace{3.5cm}\epsfbox{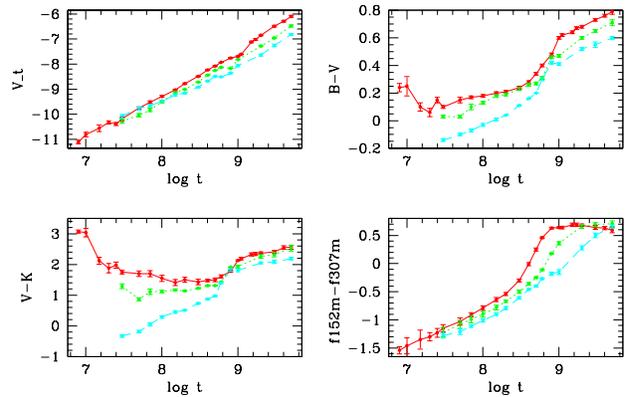} 
\vspace{-3cm}
\caption[h]{The integrated V magnitude and colours 
for three different choice of Z: Z$~=~2\cdot 10^{-2}$
({\it solid line}), 
Z$~=~6\cdot 10^{-3}$ ({\it dotted line}) and 
Z$~=~10^{-3}$ ({\it dashed line}).}
\label{fig10}
\end{figure}

The behavior of HST UV colours discloses differences in their pattern 
according to the different metallicities . As well known, the 
RGB location in the HR diagram depends on metallicity, in the sense that 
cooler RGB are expected from stellar models of larger metallicity. 
For this reason the low metallicity population synthesis models 
have hotter RGB than RF models do. As a consequence, they are less affected
by red leak and they maintain a monotonic behavior for the older ages 
presented here.

\begin{figure}[htb]
\epsfxsize=8.8cm 
\hspace{3.5cm}\epsfbox{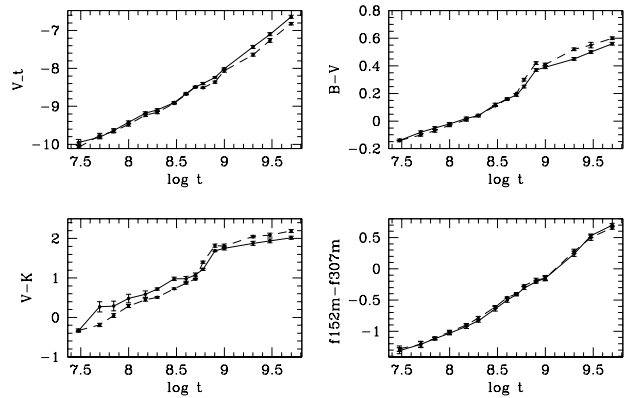} 
\vspace{-3cm}
\caption[h]{As in Fig. 10 but for Z$~=~10^{-3}$ and two
different value of the original helium content 
$Y~=~0.23$ ({\it solid line}) and 
$Y~=~0.27$ ({\it dashed line}). }
\label{fig11}
\end{figure}

Finally, Fig. 11 (Table 3--4) shows that little variations are found for
different assumptions on the Y value. In particular, variations in Y do  not
affect the HST UV colours and most of the large band colours (a small
difference can be found in V$-$K colour at intermediate ages) for the range of
ages considered in the present work.

\section{Comparison with previous models and to LMC clusters}

As a first step, Fig. 12 compares the  UV two-colours diagram  
(C(15$-$31) vs C(18$-$28)) presented in the already quoted 
paper by Barbero et al. (1990) with analogous results, but obtained
from the present computations over the relevant range of ages (t$\leq$ 1 Gyr). 
The time evolution of the diagrams appear in rather
good agreement, supporting the scenario discussed in that paper.
However, a not negligible  difference can be found in the absolute 
calibration of the age which is now revised according to Table 5. 
In the same figure we plot observational data for the sample
of LMC stellar clusters presented by Barbero et al.(1990), implemented with
more recent data by Cassatella et al. (1996). 

\begin{figure}[tb]
\epsfxsize=8.8cm 
\hspace{3.5cm}\epsfbox{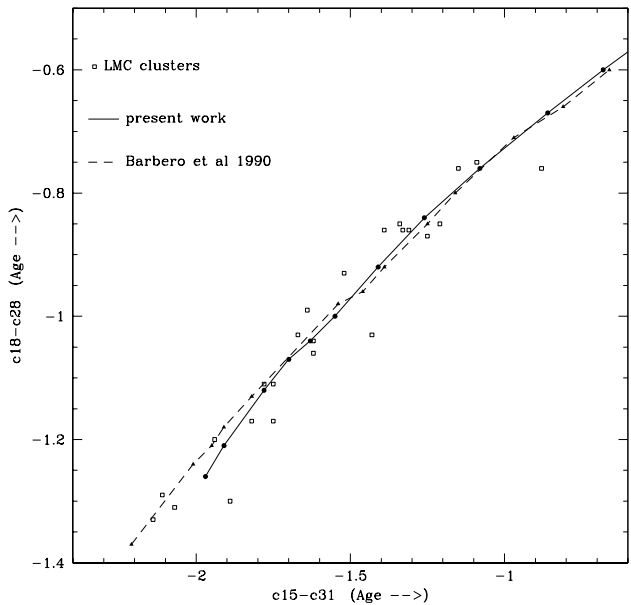} 
\caption[h]{Two UV colours diagram as in
Barbero et al 1990. ({\it dashed line})  compared to the present
work ({\it solid line}). Open square represent observational data of 
LMC clusters. }
\label{fig12}
\end{figure}

Further comparisons can be made only for broad band colours, since HST 
filters, as presented in this work, up to now have been presented only by Yi, 
Demarque \& Oemler (1995) but for older  populations ($\ge 12$ Gyr). 
One finds a general agreement with the behavior of UBVK colours
given by Bruzual \& Charlot (1993) for simple stellar populations. 
These authors adopted stellar evolution tracks by Maeder \& Meynet 
(1989, 1991) in which a rather efficient core overshooting is assumed,
thus predicting  different (larger) evolutionary times. 
Our models predict the occurrence of red supergiants in moderately
metal poor massive stars, not found by Maeder \& Meynet  because of
the adoption of the Schwarzschild convective criterion 
(see, for a discussion on that 
matter Stothers \& Chin 1992, Brocato \& Castellani 1993 and references
therein). Correspondingly, we predict redder
colour for very young, moderately metal poor clusters.

To enter in more details, let us compare our results with 
the colours more recently presented by Bressan et al. (1994: B94)
and by Bruzual and Charlot (GISSEL95 and GISSEL96  : Leitherer et al. 1996),
again on the basis of stellar evolutionary models allowing for an
efficient core overshooting. The comparison, as  given
in Fig. 13 for clusters with solar metallicity, discloses a 
remarkable agreement in both U$-$B and B$-$V colours. As expected, for a given
U$-$B colour B94 and GISSEL95/96 give larger ages. 
As a matter of fact, this colour is
dominated by the luminous termination of the cluster MS, and overshooting 
gives (roughly) a similar termination but for larger ages than 
canonical computations do. This difference vanishes for the larger ages,
perhaps -- at least in part -- because for less massive stars in the range
1.0 to 1.5 M$_{\odot}$ B94 adopted a reduced amount of overshooting.

\begin{figure}[t]
\epsfxsize=8.8cm 
\hspace{3.5cm}\epsfbox{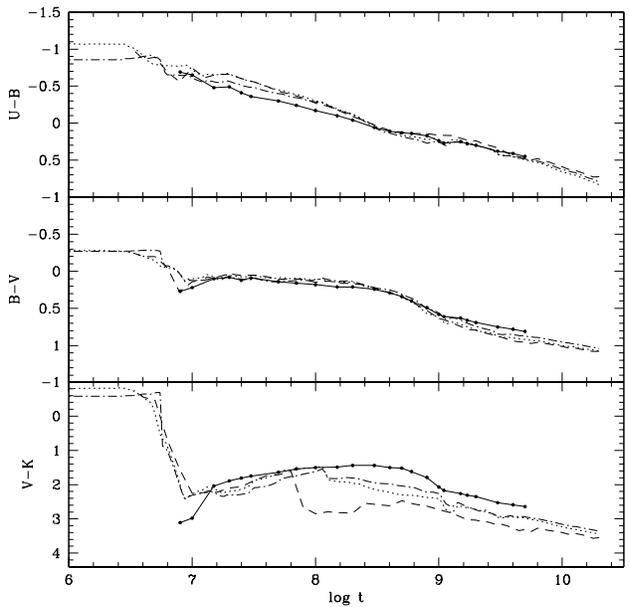} 
\caption[h]{Present  integrated colours ({\it solid line}) 
compared to similar models by Bressan et al. (1994) ({\it dashed line}),
Bruzual \& Charlot GISSEL95 ({\it dot-dashed line}) and 
GISSEL96 ({\it dotted line}).}
\label{fig13}
\end{figure}

The comparison of B$-$V colours deserves a bit more discussion.
One finds a close similarity of results for log t $\leq$ 9, 
whereas for larger ages B94 predicts redder colours. This last 
occurrence can be taken as an evidence that for old cluster
dominated by Red Giants B94 gives cooler Giant Branches than we do.
Both computations use a mixing length l$\simeq$ 1.6 H$_{p}$, and the above
occurrence should be likely ascribed to the use by B94 of improved model 
atmosphere by Kurucz (1992) which give slightly redder colour
for red giants. However, we have already discussed in the introduction the
evidence that even our branches appear too red in comparison 
with actual clusters. Thus we can only conclude that
predictions about red colour indexes should wait for
evolutionary computations calibrated on the cluster 
giant branch rather than on the Sun. GISSEL95/96 are slightly 
bluer than B94 probably due to the different assumption on the 
model atmospheres, however more details can be found in Charlot, 
Worthey \& Bressan (1996).

Finally, one finds that our, B94 and GISSEL95/96 V$-$K colours appear
reasonably similar only in a restricted range of ages, namely
for 7 $<$log t $<$ 8. For smaller age we predict
more red giants and, thus, redder colours than B94 does. 
A possible explanation could be related to the fact that
B94 models in this range of masses have been possibly
interpolated between the 12 $M_{\odot}$ (which has a He-burning loop 
in the red side of the CMD) and the 30 $M_{\odot}$(which has the 
He-burning phase at high temperature,
i.e. low emission in the K band). Consequently
their V$-$K colours move to the blue following this interpolation.

The discrepancy at the larger age can be understood bearing in mind that
this colour largely follows the occurrence of AGB stars (see, e.g.,
Ferraro et al. 1994). The results by B94 show
a jump in the V$-$K colour at $\log t~=~7.8$
due to the Phase Transition (Renzini \& Buzzoni 1986) at $t(M_{UP})$. 
In our models we included the AGB phase when $M \le t(M_{UP})$ 
but neglecting the thermal pulses phase (TP-AGB). 
GISSEL95/96 models present a trend similar to our model even if they are
sistematically redder than our of about 0.5 mag. This discrepancy is
again related to the different treatment of the TP-AGB phase which foresee
a different number of expected TP-AGB stars (see also Charlot, 
Worthey \& Bressan 1996).
However, we recall that 
the actual time extension of the TP-AGB is still a debated 
question (Bl\"{o}cker \& Sch\"{o}nberner 1991, Renzini 1992).

\begin{figure}[htb]
\epsfxsize=8cm 
\hspace{3.5cm}\epsfbox{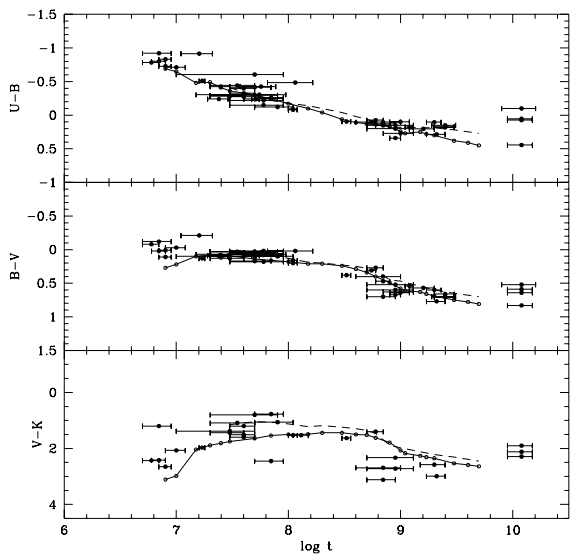} 
\caption[h]{Present integrated colours (Z$~=~$0.02 {\it solid line},
Z$~=~$$6 \cdot 10^{-3}$ {\it dashed line})
compared to observational data for MC clusters. Error bars on cluster 
age determination are from the literature.}
\label{fig14}
\end{figure}

To shed light on this problem and, more generally, to test 
theoretical predictions Fig. 14 compares theoretical prediction 
with observational U$-$B, B$-$V and V$-$K colours vs age relations for 
MC clusters with known age from isochrone fitting. The population 
synthesis models for each colour are  plotted  for both 
Z$~=~$0.02 and Z$~=~$0.006. 
The agreement is good for U$-$B and B$-$V, and the somewhat
large uncertainties on age do not allow any discrimination about
the efficiency of overshooting. The V$-$K colour is much less clear, and 
-- in particular -- one finds
clusters distributed on both the alternative predictions 
of curves in Fig. 14. The possible statistical fluctuations, 
which for V$-$K colour can be larger than $\sigma \simeq 0.9$ mag for clusters
with $M_V \ge -7$, and the uncertainties in the age evaluation of MC 
stellar clusters could largely take into account  the scatter of 
observational data, forbidding any conclusion on that matter.

Beyond such an uncertainty, let us notice that the fair 
agreement between our broad band U$-$B and B$-$V colours
and those by B94 and by Bruzual \& Charlot (GISSEL95/96) 
suggest
that the results of population synthesis appear rather solid, given current
uncertainties in the stellar evolution theories.
\par

\section{Discussion and Conclusions}

To further explore the capability of HST filters to give information
on the cluster ages Fig. 15 compares  the (F152M$-$F307M~vs~F170M$-$F278M) 
two colour diagram for the RF model with a similar diagram but for filters
without red-leak. One finds that for F152M-F307M 
values larger than $-$0.2 the red leak effect dominates,
producing the small hook at F152M-F307M$\simeq 0.7$. This is
the consequence of the fact, already shown in Fig. 7, that at F152M$-$F307M 
$\simeq 0.5$ HST colours 
saturate and their relation with the age is not reliable any more.
It follows that HST filters can give cluster ages only 
for blue colours (F152M$-$F307M $\le -0.2$), i.e., for ages smaller
than 1~Gyr. As shown in the same figure, this should not be the case 
for the C(15$-$31)~vs~C(18$-$28) colours, which are not affected by any
red leak.

\begin{figure}[htb]
\epsfxsize=8.8cm 
\hspace{3.5cm}\epsfbox{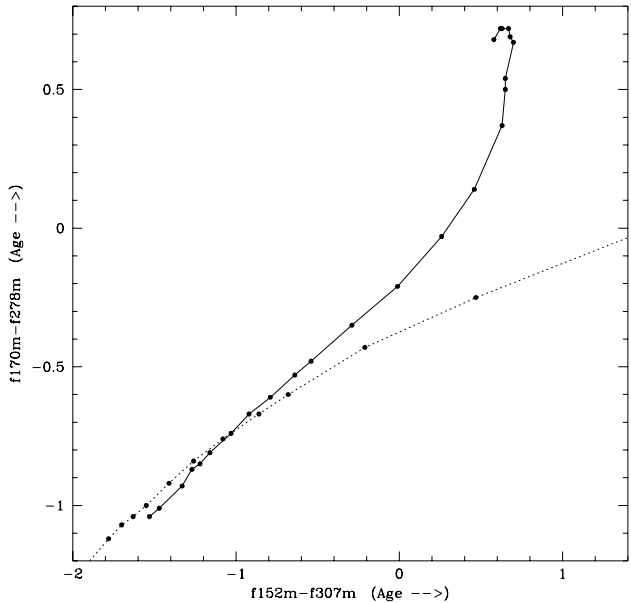} 
\caption[h]{The UV two colour diagram for HST filters 
for the RF models ({\it solid line})compared with the not red-leaked
c(15$-$31) vs c(18$-$28) relation ({\it dotted line}) (see text). }
\label{fig15}
\end{figure}

The calibration of the two colours relation in terms of cluster ages
can be derived from Table 6 for selected metallicity values.  As
discussed in section 4, the effects of the Y or IMF appear of minor
relevance. A major uncertainty in the evaluation of the age is due to
the metallicity, in the sense that for a given colour a low
metallicity population would have older ages.  However, we have
already found that V$-$K colours appear quite sensitive to the cluster
metallicity. We suggest that UV integrated colours together with V$-$K
colours could be important tools to infer information on the age {\it
and} metallicity of relatively young population in distant stellar
systems.  Although the present theoretical uncertainties shown in
Fig. 13 still require precise and conclusive comparison between
stellar evolutionary models of massive stars and young stellar
clusters, we present two cases as an example of the potential
application of firm theoretical constraints on integrated cluster
colours.  As a first approach to this problem, let us discuss HST
observations of suspected young clusters in the field of the galaxies
NGC 3921 (Schweizer et al. 1996) and NGC 1275 (Holtzman et
al. 1992). The lower panels in Fig. 16 show the observed distribution
of V$-$I colour for clusters in NGC 3921 (left panel) and of V$-$R
colours for clusters in NGC 1275. The upper panels show the predicted
time evolution of the corresponding colours (age on y axis) of three
simple stellar populations with the labeled metallicities.

\begin{figure}[htb]
\epsfxsize=8.8cm 
\hspace{3.5cm}\epsfbox{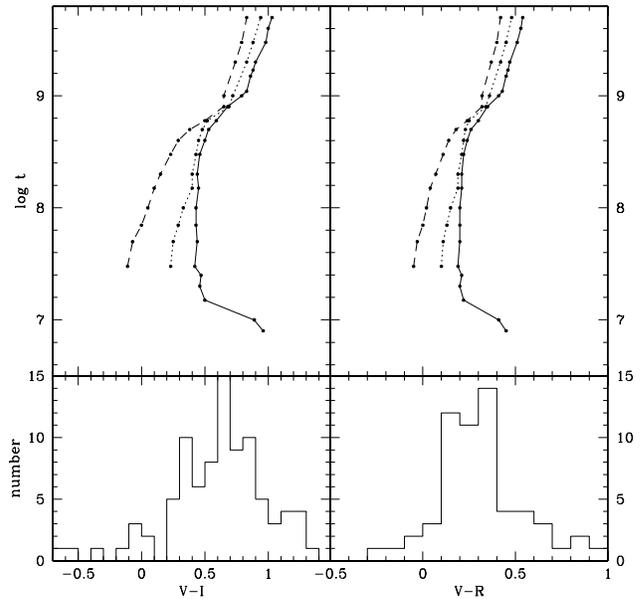} 
\caption[h]{Observed colour distribution of cluster in NGC 3921 ({\it lower-left panel}) 
and NGC 1275 ({\it lower-right panel}). The upper panels report the theoretical 
integrated V$-$I ({\it upper-left panel}) and V$-$R ({\it upper-right panel}) 
colours as function of the age. The three evolutive lines represent the different 
metallicity (Z$~=~2 \cdot 10^{-2}$ ({\it solid line}), Z$~=~6 \cdot 10^{-3}$
 ({\it dotted line}) and Z$~=~10^{-3}$ ({\it dashed line}).}
\label{fig16}
\end{figure}

By adopting the present integrated colours prediction, one would
derive that in NGC 3921 the bulk of clusters should be as old as, at
least, $t \simeq 10^9$ years (or, alternatively as young as $t \simeq
10^7$ years if a solar metallicity is assumed). However, the
occurrence of clusters at V$-$I smaller than 0.2 mag should be
interpreted as an evidence for these clusters being both much younger
and metal poor. Similar consideration can be done for clusters in NGC
1275 V$-$R$\le$ 0.2. This colour is never reached by solar metallicity
clusters so these stellar systems should have a lower metallicity and
an age not exceeding $t \simeq 4~\cdot 10^8$years.

As a conclusion, we recall that in this paper 
we have presented theoretical HST and broad band colours from  
population synthesis models  based on a homogeneous set of stellar 
evolutionary tracks as
computed under canonical (no overshooting) assumptions, covering
the range of cluster ages  from t$~=~8$ Myr to t$~=~5$ Gyr for three 
different metallicities (Z$~=~$0.02, 0.006, and 0.001). Statistic 
fluctuations in the cluster population have been
investigated, assessing the expected standard fluctuations of the
various colours as a function of the cluster integrated V absolute
magnitude.

We show that the red leak in HST UV filters can deeply 
affect the expected fluxes and colours. However, we find that 
for F152M-F307M$\le 0.5$ and for F170M$-$F278M$\le 0.5 $ (which 
means ages lower than 1 Gyr) the HST UV colours can still be used 
to infer reliable indication on the age of distant clusters. 
Moreover, one finds that 
the age calibration of these colours is scarcely affected by 
the amount of original helium or by the  assumed IMF. On this 
basis, we present a calibration of the 
HST UV two-colours (F152M$-$F307M~vs~F170M$-$F278M) diagrams in terms of
cluster ages for the three above quoted metallicities.

Theoretical predictions for the time behavior
of broad band colours (UBVK) appear in good agreement with data 
for LMC stellar clusters with known age, though with
some open questions about the actual run of V$-$K colours.
We suggest that the combined use of HST UV colours and IR colours (V$-$K in 
particular)  should disentangle the metallicity-age effect in integrated 
colours of young stellar populations ($t \le 1 Gyr$). For this reason the 
NICMOS camera on board HST, particularly its F222M filter, is expected to be
a powerful tool in interpreting the integrated colours of distant objects in
term of age {\it and} metallicity.

An updated age calibration of C(15$-$31) and C(18$-$28) (Barbero et al. 1990)
is also presented. The comparison with IUE data of LMC stellar clusters 
shows a very good agreement.

We note that MC stellar clusters represent a very fundamental test to 
check the reliability of population synthesis models. Unfortunately, 
age and  metallicity determinations of single clusters available up 
to now are based, with few exceptions, on indirect calibrations. Thus, an
extensive observational work with the 
goal of deriving accurate and precise CMD for a large sample of MC 
stellar cluster would be of invaluable help to check the reliability 
of population synthesis results.

\begin{acknowledgements}

We would like to thank Warren Hack at STScI for his help with FOC filters and
for teaching us how to use the {\it focsim} package. This study has been
financially supported by the Italian Ministery of University, Scientific
Research and Technology (MURST) and the Italian Space Agency (ASI).   

\end{acknowledgements}

\begin{table*}
\scriptsize
\caption{a): HST colours for the RF model. 1$\sigma $ error is also reported.}
\begin{tabular}{cccccccccc}
\hline
\hline
 & & & & & & & & &   \\ 
age (Myr) &  M$_{\rm{TO}}$ & V$_{\rm{tot}}$ & F152m-V & F170m-V & F253m-V & F278m-V &  f307m-V & f480lp-V & f606w-V  \\
 & & & & & & & & &   \\  
 \hline
   & & & & & & & & &   \\ 
    8 &19.74 &-11.14$\pm $0.07 &-2.66$\pm $0.05 &-2.45$\pm $0.05 & -1.71$\pm $0.05 & -1.41$\pm $0.05 & -1.13$\pm $0.04 & -0.04$\pm $0.01 & -0.24$\pm $0.01 \\
 & & & & & & & &&   \\   
   10 &16.07 &-10.83$\pm $0.09 &-2.58$\pm $0.10 &-2.38$\pm $0.10 & -1.65$\pm $0.10 & -1.37$\pm $0.10 & -1.11$\pm $0.10 & -0.05$\pm $0.01 & -0.22$\pm $0.03\\
& & & & & & & &&   \\   
   15 &12.20 &-10.67$\pm $0.13 &-2.21$\pm $0.13 &-2.02$\pm $0.13 & -1.34$\pm $0.13 & -1.09$\pm $0.12 & -0.88$\pm $0.11 & -0.08$\pm $0.01 & -0.11$\pm $0.01\\
& & & & & & & &&   \\   
   20 &10.31 &-10.41$\pm $0.07 &-2.17$\pm $0.06 &-1.98$\pm $0.05 & -1.33$\pm $0.05 & -1.11$\pm $0.05 & -0.90$\pm $0.04 & -0.09$\pm $0.01 & -0.10$\pm $0.02\\
& & & & & & & &&   \\
   25 & 9.08 &-10.33$\pm $0.06 &-1.95$\pm $0.06 &-1.76$\pm $0.06 & -1.12$\pm $0.06 & -0.91$\pm $0.06 & -0.73$\pm $0.05 & -0.08$\pm $0.00 & -0.11$\pm $0.01\\
& & & & & & & &&   \\
   30 & 8.22 &-10.18$\pm $0.07 &-1.80$\pm $0.05 &-1.62$\pm $0.05 & -0.99$\pm $0.05 & -0.81$\pm $0.05 & -0.64$\pm $0.04 & -0.09$\pm $0.00 & -0.09$\pm $0.01\\
& & & & & & & &&   \\
   50 & 6.42 & -9.77$\pm $0.05 &-1.53$\pm $0.05 &-1.37$\pm $0.05 & -0.77$\pm $0.05 & -0.63$\pm $0.05 & -0.50$\pm $0.04 & -0.08$\pm $0.01 & -0.10$\pm $0.01\\
& & & & & & & &&    \\
   70 & 5.52 & -9.59$\pm $0.05 &-1.27$\pm $0.03 &-1.12$\pm $0.03 & -0.55$\pm $0.03 & -0.45$\pm $0.03 & -0.35$\pm $0.02 & -0.08$\pm $0.00 & -0.10$\pm $0.01\\
& & & & & & & &&   \\
  100 & 4.73 & -9.37$\pm $0.04 &-0.97$\pm $0.03 &-0.84$\pm $0.03 & -0.30$\pm $0.03 & -0.23$\pm $0.03 & -0.18$\pm $0.03 & -0.07$\pm $0.01 & -0.10$\pm $0.01\\
& & & & & & & & &   \\
  150 & 4.01 & -9.10$\pm $0.03 &-0.62$\pm $0.03 &-0.51$\pm $0.03 & -0.01$\pm $0.03 &  0.02$\pm $0.03 &  0.02$\pm $0.02 & -0.07$\pm $0.00 & -0.11$\pm $0.01\\
& & & & & & & &  &  \\
  200 & 3.60 & -8.86$\pm $0.03 &-0.40$\pm $0.02 &-0.31$\pm $0.02 &  0.16$\pm $0.02 &  0.17$\pm $0.02 &  0.14$\pm $0.02 & -0.07$\pm $0.00 & -0.11$\pm $0.01\\
& & & & & & & & &   \\
  300 & 3.09 & -8.57$\pm $0.02 & 0.06$\pm $0.02 & 0.10$\pm $0.02 &  0.51$\pm $0.02 &  0.45$\pm $0.02 &  0.35$\pm $0.02 & -0.07$\pm $0.00 & -0.11$\pm $0.00\\
& & & & & & & & &   \\
  400 & 2.78 & -8.35$\pm $0.02 & 0.49$\pm $0.02 & 0.45$\pm $0.02 &  0.78$\pm $0.02 &  0.66$\pm $0.02 &  0.50$\pm $0.02 & -0.06$\pm $0.00 & -0.12$\pm $0.01\\
& & & & & & & & &   \\
  500 & 2.56 & -8.19$\pm $0.02 & 0.85$\pm $0.02 & 0.78$\pm $0.02 &  1.00$\pm $0.02 &  0.81$\pm $0.02 &  0.59$\pm $0.02 & -0.05$\pm $0.00 & -0.14$\pm $0.01\\
& & & & & & & &  &  \\
  600 & 2.40 & -8.07$\pm $0.01 & 1.13$\pm $0.01 & 1.08$\pm $0.01 &  1.19$\pm $0.01 &  0.94$\pm $0.01 &  0.67$\pm $0.01 & -0.04$\pm $0.00 & -0.15$\pm $0.00\\
& & & & & & & &  &  \\
  800 & 2.16 & -7.88$\pm $0.02 & 1.44$\pm $0.01 & 1.54$\pm $0.02 &  1.52$\pm $0.02 &  1.17$\pm $0.02 &  0.81$\pm $0.01 & -0.03$\pm $0.00 & -0.18$\pm $0.00\\
& & & & & & & & &   \\
 1000 & 2.00 & -7.82$\pm $0.02 & 1.63$\pm $0.01 & 1.93$\pm $0.02 &  1.87$\pm $0.02 &  1.43$\pm $0.02 &  0.98$\pm $0.01 & -0.01$\pm $0.01 & -0.22$\pm $0.00\\
& & & & & & & & &   \\
 1100 & 1.93 & -7.76$\pm $0.02 & 1.69$\pm $0.01 & 2.06$\pm $0.02 &  2.00$\pm $0.02 &  1.52$\pm $0.02 &  1.04$\pm $0.02 & -0.01$\pm $0.00 & -0.23$\pm $0.00\\
& & & & & & & & &   \\
 1500 & 1.74 & -7.36$\pm $0.03 & 1.74$\pm $0.02 & 2.25$\pm $0.02 &  2.17$\pm $0.03 &  1.58$\pm $0.03 &  1.04$\pm $0.02 & -0.01$\pm $0.00 & -0.24$\pm $0.00\\
& & & & & & & &  &  \\
 1700 & 1.67 & -7.29$\pm $0.03 & 1.79$\pm $0.02 & 2.38$\pm $0.02 &  2.32$\pm $0.03 &  1.69$\pm $0.03 &  1.11$\pm $0.02 & -0.00$\pm $0.00 & -0.25$\pm $0.00\\
& & & & & & & & &   \\
 2000 & 1.58 & -7.12$\pm $0.02 & 1.84$\pm $0.01 & 2.51$\pm $0.01 &  2.48$\pm $0.02 &  1.79$\pm $0.02 &  1.17$\pm $0.02 &  0.00$\pm $0.00 & -0.25$\pm $0.00\\
& & & & & & & &  &  \\
 3000 & 1.38 & -6.89$\pm $0.03 & 1.95$\pm $0.02 & 2.75$\pm $0.02 &  2.81$\pm $0.03 &  2.03$\pm $0.03 &  1.32$\pm $0.03 &  0.01$\pm $0.00 & -0.27$\pm $0.01\\
& & & & & & & & &   \\
 4000 & 1.27 & -6.68$\pm $0.03 & 2.00$\pm $0.02 & 2.85$\pm $0.02 &  2.95$\pm $0.03 &  2.13$\pm $0.03 &  1.38$\pm $0.02 &  0.02$\pm $0.00 & -0.28$\pm $0.01\\
& & & & & & & & &   \\
 5000 & 1.18 & -6.55$\pm $0.04 & 2.04$\pm $0.02 & 2.92$\pm $0.03 &  3.09$\pm $0.04 &  2.24$\pm $0.04 &  1.46$\pm $0.04 &  0.02$\pm $0.00 & -0.29$\pm $0.01\\
& & & & & & & & &   \\
\hline
\hline
\end{tabular}
\end{table*}

\addtocounter{table}{-1}
\begin{table*}
\scriptsize
\caption{b): Broad band colours for the RF model.}
\begin{tabular}{ccccccccc}
\hline
\hline
& & & & & & & &   \\ 
age (Myr) &  V$_{\rm{tot}}$ & U-V &  B-V & V-R & V-I & V-J & V-K & V-L \\
 & & & & & & & &   \\  
 \hline
   & & & & & & & &   \\ 
    8 &-11.14 & -0.42 $\pm $0.04 &  0.27 $\pm $0.03 &  0.45 $\pm $0.02 &  0.96 $\pm $0.04 &  2.01 $\pm $0.05 &  3.11 $\pm $0.05 &  3.16 $\pm $0.05\\
& & & & & & & &   \\ 
   10 &-10.83 & -0.43 $\pm $0.10 &  0.22 $\pm $0.07 &  0.41 $\pm $0.05 &  0.89 $\pm $0.09 &  1.90 $\pm $0.13 &  2.98 $\pm $0.14 &  3.03 $\pm $0.15\\
& & & & & & & &   \\ 
   15 &-10.67 & -0.38 $\pm $0.07 &  0.10 $\pm $0.03 &  0.22 $\pm $0.02 &  0.50 $\pm $0.05 &  1.19 $\pm $0.09 &  2.04 $\pm $0.11 &  2.09 $\pm $0.21\\
& & & & & & & &   \\ 
   20 &-10.41 & -0.41 $\pm $0.03 &  0.08 $\pm $0.03 &  0.20 $\pm $0.03 &  0.46 $\pm $0.06 &  1.10 $\pm $0.12 &  1.89 $\pm $0.15 &  1.93 $\pm $0.15\\
& & & & & & & &   \\ 
   25 &-10.33 & -0.29 $\pm $0.03 &  0.12 $\pm $0.02 &  0.21 $\pm $0.02 &  0.47 $\pm $0.04 &  1.08 $\pm $0.07 &  1.81 $\pm $0.10 &  1.85 $\pm $0.10\\
& & & & & & & &   \\ 
   30 &-10.18 & -0.27 $\pm $0.02 &  0.09 $\pm $0.01 &  0.19 $\pm $0.01 &  0.42 $\pm $0.02 &  1.00 $\pm $0.05 &  1.75 $\pm $0.06 &  1.78 $\pm $0.07\\
& & & & & & & &   \\ 
   50 & -9.77 & -0.16 $\pm $0.03 &  0.14 $\pm $0.02 &  0.20 $\pm $0.02 &  0.44 $\pm $0.03 &  0.98 $\pm $0.06 &  1.64 $\pm $0.09 &  1.68 $\pm $0.09\\
& & & & & & & &   \\ 
   70 & -9.59 & -0.08 $\pm $0.02 &  0.16 $\pm $0.01 &  0.20 $\pm $0.01 &  0.43 $\pm $0.03 &  0.94 $\pm $0.06 &  1.54 $\pm $0.10 &  1.57 $\pm $0.10\\
& & & & & & & &   \\ 
  100 & -9.37 &  0.01 $\pm $0.02 &  0.18 $\pm $0.01 &  0.20 $\pm $0.01 &  0.43 $\pm $0.03 &  0.92 $\pm $0.06 &  1.50 $\pm $0.10 &  1.53 $\pm $0.10\\
& & & & & & & &   \\ 
  150 & -9.10 &  0.11 $\pm $0.02 &  0.21 $\pm $0.01 &  0.21 $\pm $0.01 &  0.45 $\pm $0.02 &  0.92 $\pm $0.06 &  1.49 $\pm $0.10 &  1.51 $\pm $0.10\\
& & & & & & & &   \\ 
  200 & -8.86 &  0.17 $\pm $0.01 &  0.21 $\pm $0.01 &  0.21 $\pm $0.01 &  0.44 $\pm $0.02 &  0.91 $\pm $0.05 &  1.44 $\pm $0.08 &  1.47 $\pm $0.08\\
& & & & & & & &   \\ 
  300 & -8.57 &  0.30 $\pm $0.01 &  0.24 $\pm $0.01 &  0.22 $\pm $0.01 &  0.46 $\pm $0.02 &  0.92 $\pm $0.05 &  1.44 $\pm $0.09 &  1.47 $\pm $0.10\\
& & & & & & & &   \\ 
  400 & -8.35 &  0.40 $\pm $0.01 &  0.29 $\pm $0.01 &  0.24 $\pm $0.01 &  0.50 $\pm $0.01 &  0.98 $\pm $0.03 &  1.50 $\pm $0.04 &  1.52 $\pm $0.05\\
& & & & & & & &   \\ 
  500 & -8.19 &  0.47 $\pm $0.02 &  0.34 $\pm $0.01 &  0.26 $\pm $0.01 &  0.53 $\pm $0.02 &  1.01 $\pm $0.04 &  1.52 $\pm $0.06 &  1.54 $\pm $0.06\\
& & & & & & & &   \\ 
  600 & -8.07 &  0.54 $\pm $0.01 &  0.40 $\pm $0.01 &  0.30 $\pm $0.01 &  0.59 $\pm $0.01 &  1.10 $\pm $0.03 &  1.62 $\pm $0.05 &  1.65 $\pm $0.05\\
& & & & & & & &   \\ 
  800 & -7.88 &  0.66 $\pm $0.02 &  0.49 $\pm $0.01 &  0.35 $\pm $0.01 &  0.68 $\pm $0.01 &  1.24 $\pm $0.02 &  1.79 $\pm $0.03 &  1.82 $\pm $0.03\\
& & & & & & & &   \\ 
 1000 & -7.82 &  0.82 $\pm $0.01 &  0.58 $\pm $0.01 &  0.41 $\pm $0.01 &  0.79 $\pm $0.01 &  1.43 $\pm $0.02 &  2.07 $\pm $0.02 &  2.10 $\pm $0.02\\
& & & & & & & &   \\ 
 1100 & -7.76 &  0.88 $\pm $0.02 &  0.61 $\pm $0.01 &  0.43 $\pm $0.01 &  0.83 $\pm $0.01 &  1.51 $\pm $0.03 &  2.17 $\pm $0.04 &  2.21 $\pm $0.04\\
& & & & & & & &   \\ 
 1500 & -7.36 &  0.88 $\pm $0.02 &  0.63 $\pm $0.01 &  0.45 $\pm $0.01 &  0.86 $\pm $0.02 &  1.55 $\pm $0.03 &  2.26 $\pm $0.05 &  2.29 $\pm $0.05\\
& & & & & & & &   \\ 
 1700 & -7.29 &  0.94 $\pm $0.02 &  0.66 $\pm $0.01 &  0.46 $\pm $0.01 &  0.88 $\pm $0.01 &  1.60 $\pm $0.03 &  2.31 $\pm $0.05 &  2.35 $\pm $0.05\\
& & & & & & & &   \\ 
 2000 & -7.12 &  0.99 $\pm $0.01 &  0.69 $\pm $0.01 &  0.47 $\pm $0.00 &  0.90 $\pm $0.01 &  1.63 $\pm $0.02 &  2.35 $\pm $0.04 &  2.38 $\pm $0.04\\
& & & & & & & &   \\ 
 3000 & -6.89 &  1.13 $\pm $0.03 &  0.75 $\pm $0.01 &  0.51 $\pm $0.01 &  0.98 $\pm $0.02 &  1.76 $\pm $0.03 &  2.53 $\pm $0.05 &  2.57 $\pm $0.06\\
& & & & & & & &   \\ 
 4000 & -6.68 &  1.19 $\pm $0.02 &  0.78 $\pm $0.01 &  0.53 $\pm $0.01 &  1.00 $\pm $0.02 &  1.80 $\pm $0.04 &  2.59 $\pm $0.06 &  2.63 $\pm $0.06\\
& & & & & & & &   \\ 
 5000 & -6.55 &  1.26 $\pm $0.04 &  0.81 $\pm $0.02 &  0.54 $\pm $0.01 &  1.03 $\pm $0.02 &  1.84 $\pm $0.05 &  2.64 $\pm $0.07 &  2.68 $\pm $0.07\\
& & & & & & & &  \\
\hline
\hline
\end{tabular}
\end{table*}

\begin{table*}
\scriptsize
\caption{a): HST colours for the Y=0.27, Z=$6\cdot 10^{-3}$ model.}
\begin{tabular}{cccccccccc}
\hline
\hline
& & & & & & & & &  \\ 
age (Myr) & M$_{\rm{TO}}$ & V$_{\rm{tot}}$ & F152m-V & F170m-V & F253m-V & F278m-V & f307m-V & f480lp-V & f606w-V \\
& & & & & & & & &  \\ 
   30 & 7.91 & -10.33 $\pm $0.08 &  -1.96 $\pm $0.07 &  -1.77 $\pm $0.07 &  -1.13 $\pm $0.06 &  -0.94 $\pm $0.06 &  -0.76 $\pm $0.05 &  -0.10 $\pm $0.01 &  -0.05 $\pm $0.01\\
& & & & & & & & &  \\
   50 & 6.05 & -10.06 $\pm $0.07 &  -1.57 $\pm $0.06 &  -1.39 $\pm $0.06 &  -0.78 $\pm $0.05 &  -0.63 $\pm $0.05 &  -0.49 $\pm $0.04 &  -0.10 $\pm $0.00 &  -0.05 $\pm $0.00\\
& & & & & & & & &  \\
   70 & 5.15 &  -9.83 $\pm $0.07 &  -1.34 $\pm $0.05 &  -1.17 $\pm $0.05 &  -0.59 $\pm $0.05 &  -0.45 $\pm $0.05 &  -0.34 $\pm $0.04 &  -0.09 $\pm $0.00 &  -0.06 $\pm $0.01\\
& & & & & & & & &  \\
  100 & 4.39 &  -9.54 $\pm $0.04 &  -1.14 $\pm $0.04 &  -0.99 $\pm $0.04 &  -0.43 $\pm $0.04 &  -0.32 $\pm $0.04 &  -0.24 $\pm $0.03 &  -0.09 $\pm $0.00 &  -0.08 $\pm $0.00\\
& & & & & & & & &  \\
  150 & 3.69 &  -9.24 $\pm $0.04 &  -0.88 $\pm $0.03 &  -0.74 $\pm $0.03 &  -0.22 $\pm $0.03 &  -0.15 $\pm $0.03 &  -0.10 $\pm $0.03 &  -0.08 $\pm $0.01 &  -0.10 $\pm $0.01\\
& & & & & & & & &  \\
  200 & 3.27 &  -9.06 $\pm $0.02 &  -0.66 $\pm $0.02 &  -0.52 $\pm $0.01 &  -0.04 $\pm $0.01 &   0.00 $\pm $0.01 &   0.01 $\pm $0.01 &  -0.07 $\pm $0.00 &  -0.10 $\pm $0.00\\
& & & & & & & & &  \\
  300 & 2.77 &  -8.76 $\pm $0.01 &  -0.34 $\pm $0.02 &  -0.22 $\pm $0.02 &   0.22 $\pm $0.02 &   0.21 $\pm $0.02 &   0.17 $\pm $0.02 &  -0.07 $\pm $0.00 &  -0.10 $\pm $0.01\\
& & & & & & & & &  \\
  400 & 2.47 &  -8.53 $\pm $0.01 &  -0.08 $\pm $0.02 &   0.02 $\pm $0.02 &   0.41 $\pm $0.02 &   0.37 $\pm $0.02 &   0.29 $\pm $0.01 &  -0.06 $\pm $0.01 &  -0.11 $\pm $0.00\\
& & & & & & & & &  \\
  500 & 2.27 &  -8.37 $\pm $0.01 &   0.15 $\pm $0.01 &   0.23 $\pm $0.01 &   0.58 $\pm $0.02 &   0.51 $\pm $0.01 &   0.39 $\pm $0.01 &  -0.06 $\pm $0.00 &  -0.12 $\pm $0.00\\
& & & & & & & & &  \\
  600 & 2.12 &  -8.31 $\pm $0.02 &   0.38 $\pm $0.01 &   0.43 $\pm $0.01 &   0.75 $\pm $0.01 &   0.64 $\pm $0.01 &   0.48 $\pm $0.01 &  -0.05 $\pm $0.00 &  -0.13 $\pm $0.00\\
& & & & & & & & &  \\
  800 & 1.90 &  -8.31 $\pm $0.01 &   0.93 $\pm $0.01 &   0.96 $\pm $0.01 &   1.19 $\pm $0.01 &   1.00 $\pm $0.01 &   0.75 $\pm $0.00 &  -0.03 $\pm $0.00 &  -0.18 $\pm $0.00\\
& & & & & & & & &  \\
 1000 & 1.75 &  -8.00 $\pm $0.02 &   1.12 $\pm $0.02 &   1.11 $\pm $0.02 &   1.25 $\pm $0.02 &   1.03 $\pm $0.02 & 0.77 $\pm $0.02 &  -0.03 $\pm $0.00 &  -0.18 $\pm $0.00\\
& & & & & & & & &  \\
 2000 & 1.38 &  -7.47 $\pm $0.03 &   1.58 $\pm $0.02 &   1.79 $\pm $0.03 &   1.71 $\pm $0.04 &   1.31 $\pm $0.03  & 0.92 $\pm $0.03 &  -0.01 $\pm $0.00 &  -0.22 $\pm $0.00\\
& & & & & & & & &  \\
 3000 & 1.20 &  -7.29 $\pm $0.03 &   1.71 $\pm $0.02 &   2.10 $\pm $0.03 &   1.98 $\pm $0.03 &   1.47 $\pm $0.03 &   1.00 $\pm $0.03 &  -0.00 $\pm $0.00 &  -0.24 $\pm $0.01\\
& & & & & & & & &  \\
 5000 & 1.05 &  -7.03 $\pm $0.04 &   1.82 $\pm $0.03 &   2.41 $\pm $0.03 &   2.28 $\pm $0.04 &   1.66 $\pm $0.04 &   1.11 $\pm $0.04  & 0.01 $\pm $0.01 &  -0.26 $\pm $0.01\\
& & & & & & & & &  \\
\hline
\hline
\end{tabular}
\end{table*}

\addtocounter{table}{-1}
\begin{table*}
\scriptsize
\caption{b): Broad band colours for the Y=0.27, Z=$6\cdot 10^{-3}$ model.}
\begin{tabular}{ccccccccc}
\hline
\hline
& & & & & & & &   \\ 
age (Myr) & V$_{\rm{tot}}$ & U-V & B-V & V-R & V-I & V-J & V-K & V-L \\
& & & & & & & &   \\ 
   30& -10.33&  -0.40$\pm $0.02&   0.01$\pm $0.01&   0.10$\pm $0.01&   0.23$\pm $0.02&   0.60$\pm $0.05&   1.15$\pm $0.08&   1.18$\pm $0.08\\
& & & & & & & &   \\  
  50& -10.06&  -0.24$\pm $0.02&   0.05$\pm $0.01&   0.11$\pm $0.01&   0.25$\pm $0.02&   0.58$\pm $0.04&   1.04$\pm $0.05&   1.06$\pm $0.06\\
& & & & & & & &   \\ 
  70&  -9.83&  -0.13$\pm $0.03&   0.09$\pm $0.02&   0.13$\pm $0.01&   0.29$\pm $0.03&   0.63$\pm $0.06&   1.06$\pm $0.09&   1.09$\pm $0.09\\
& & & & & & & &   \\ 
  100&  -9.54&  -0.05$\pm $0.02&   0.13$\pm $0.01&   0.15$\pm $0.00&   0.33$\pm $0.01&   0.69$\pm $0.03&   1.12$\pm $0.04&   1.14$\pm $0.04\\
& & & & & & & &   \\ 
  150&  -9.24&   0.05$\pm $0.02&   0.19$\pm $0.01&   0.19$\pm $0.01&   0.40$\pm $0.01&   0.79$\pm $0.02&   1.22$\pm $0.03&   1.24$\pm $0.03\\
& & & & & & & &   \\ 
  200&  -9.06&   0.10$\pm $0.01&   0.20$\pm $0.01&   0.19$\pm $0.01&   0.40$\pm $0.01&   0.79$\pm $0.01&   1.19$\pm $0.02&   1.21$\pm $0.02\\
& & & & & & & &   \\ 
  300&  -8.76&   0.19$\pm $0.01&   0.23$\pm $0.01&   0.21$\pm $0.01&   0.43$\pm $0.01&   0.83$\pm $0.02&   1.24$\pm $0.02&   1.25$\pm $0.02\\
& & & & & & & &   \\ 
  400&  -8.53&   0.27$\pm $0.01&   0.25$\pm $0.01&   0.22$\pm $0.00&   0.45$\pm $0.01&   0.87$\pm $0.01&   1.29$\pm $0.02&   1.31$\pm $0.02\\
& & & & & & & &   \\ 
  500&  -8.37&   0.34$\pm $0.01&   0.28$\pm $0.01&   0.23$\pm $0.00&   0.48$\pm $0.01&   0.92$\pm $0.01&   1.36$\pm $0.01&   1.38$\pm $0.01\\
& & & & & & & &   \\ 
  600&  -8.31&   0.41$\pm $0.01&   0.31$\pm $0.01&   0.25$\pm $0.00&   0.52$\pm $0.01&   0.99$\pm $0.01&   1.46$\pm $0.01&   1.48$\pm $0.01\\
& & & & & & & &   \\ 
  800&  -8.31&   0.63$\pm $0.01&   0.45$\pm $0.01&   0.34$\pm $0.00&   0.69$\pm $0.00&   1.29$\pm $0.01&   1.89$\pm $0.01&   1.92$\pm $0.01\\
& & & & & & & &   \\ 
 1000&  -8.00&   0.65$\pm $0.02&   0.47$\pm $0.01&   0.36$\pm $0.01&   0.72$\pm $0.02&   1.35$\pm $0.03&   1.98$\pm $0.05& 2.01$\pm $0.05\\
& & & & & & & &   \\ 
 2000&  -7.47&   0.79$\pm $0.02&   0.60$\pm $0.01&   0.42$\pm $0.01&   0.83$\pm $0.02&   1.53$\pm $0.04&   2.22$\pm $0.06&   2.25$\pm $0.06\\
& & & & & & & &   \\ 
 3000&  -7.29&   0.87$\pm $0.03&   0.65$\pm $0.01&   0.45$\pm $0.01&   0.88$\pm $0.02&   1.60$\pm $0.05&   2.31$\pm $0.07&   2.35$\pm $0.07\\
& & & & & & & &   \\ 
 5000&  -7.03&   0.97$\pm $0.04&   0.70$\pm $0.02&   0.48$\pm $0.01&   0.94$\pm $0.03&   1.70$\pm $0.06&   2.45$\pm $0.09&   2.49$\pm $0.09\\
& & & & & & & &   \\ 
\hline
\hline
\end{tabular}
\end{table*}


\begin{table*}
\scriptsize
\caption{a): HST colours for the Y=0.27, Z=$10^{-3}$ model.}
\begin{tabular}{cccccccccc}
\hline
\hline
& & & & & & & & &  \\ 
age (Myr) & M$_{\rm{TO}}$ & V$_{\rm{tot}}$ & F152m-V & F170m-V & F253m-V & F278m-V & f307m-V & f480lp-V & f606w-V \\
& & & & & & & & &  \\ 
   30&  7.61& -10.04$\pm $0.07&  -2.53$\pm $0.04&  -2.34$\pm $0.04&  -1.68$\pm $0.03&  -1.45$\pm $0.03&  -1.23$\pm $0.02&  -0.14$\pm $0.00&   0.02$\pm $0.00\\
& & & & & & & & &  \\ 
   50&  5.78&  -9.81$\pm $0.06&  -2.19$\pm $0.04&  -2.00$\pm $0.04&  -1.37$\pm $0.04&  -1.17$\pm $0.03&  -0.98$\pm $0.03&  -0.13$\pm $0.00&   0.01$\pm $0.00\\
& & & & & & & & &  \\ 
   70&  4.89&  -9.68$\pm $0.04&  -1.90$\pm $0.02&  -1.74$\pm $0.02&  -1.13$\pm $0.02&  -0.95$\pm $0.02&  -0.79$\pm $0.01&  -0.12$\pm $0.00&  -0.00$\pm $0.00\\
& & & & & & & & &  \\ 
  100&  4.13&  -9.52$\pm $0.04&  -1.62$\pm $0.02&  -1.46$\pm $0.02&  -0.89$\pm $0.02&  -0.74$\pm $0.02&  -0.60$\pm $0.02&  -0.12$\pm $0.00&  -0.01$\pm $0.00\\
& & & & & & & & &  \\ 
  150&  3.44&  -9.32$\pm $0.04&  -1.28$\pm $0.02&  -1.13$\pm $0.01&  -0.61$\pm $0.02&  -0.49$\pm $0.01&  -0.38$\pm $0.01&  -0.11$\pm $0.00&  -0.02$\pm $0.00\\
& & & & & & & & &  \\ 
  200&  3.02&  -9.21$\pm $0.03&  -0.99$\pm $0.02&  -0.86$\pm $0.02&  -0.37$\pm $0.02&  -0.28$\pm $0.02&  -0.20$\pm $0.02&  -0.10$\pm $0.00&  -0.03$\pm $0.00\\
& & & & & & & & &  \\ 
  300&  2.53&  -8.97$\pm $0.02&  -0.60$\pm $0.02&  -0.50$\pm $0.01&  -0.08$\pm $0.02&  -0.03$\pm $0.01&   0.00$\pm $0.01&  -0.09$\pm $0.00&  -0.05$\pm $0.00\\
& & & & & & & & &  \\ 
  400&  2.25&  -8.77$\pm $0.02&  -0.35$\pm $0.01&  -0.25$\pm $0.01&   0.12$\pm $0.01&   0.14$\pm $0.01&   0.12$\pm $0.01&  -0.08$\pm $0.00&  -0.07$\pm $0.00\\
& & & & & & & & &  \\ 
  500&  2.06&  -8.64$\pm $0.02&  -0.15$\pm $0.01&  -0.03$\pm $0.01&   0.29$\pm $0.01&   0.27$\pm $0.01&   0.22$\pm $0.01&  -0.07$\pm $0.00&  -0.09$\pm $0.00\\
& & & & & & & & &  \\ 
  600&  1.92&  -8.66$\pm $0.02&   0.08$\pm $0.01&   0.22$\pm $0.02&   0.52$\pm $0.01&   0.45$\pm $0.01&   0.35$\pm $0.01&  -0.06$\pm $0.00&  -0.12$\pm $0.00\\
& & & & & & & & &  \\ 
  800&  1.73&  -8.47$\pm $0.03&   0.35$\pm $0.02&   0.50$\pm $0.03&   0.81$\pm $0.03&   0.71$\pm $0.03&   0.55$\pm $0.02&  -0.04$\pm $0.00&  -0.16$\pm $0.01\\
& & & & & & & & &  \\ 
 1000&  1.60&  -8.20$\pm $0.04&   0.43$\pm $0.03&   0.56$\pm $0.04&   0.84$\pm $0.04&   0.74$\pm $0.03&   0.58$\pm $0.03&  -0.04$\pm $0.00&  -0.16$\pm $0.01\\
& & & & & & & & &  \\ 
 2000&  1.27&  -7.79$\pm $0.04&   1.02$\pm $0.03&   1.11$\pm $0.03&   1.21$\pm $0.03&   1.00$\pm $0.03&   0.75$\pm $0.02&  -0.03$\pm $0.00&  -0.19$\pm $0.00\\
& & & & & & & & &  \\ 
 3000&  1.15&  -7.63$\pm $0.05&   1.32$\pm $0.03&   1.41$\pm $0.05&   1.40$\pm $0.05&   1.12$\pm $0.04&   0.83$\pm $0.03&  -0.02$\pm $0.00&  -0.21$\pm $0.01\\
& & & & & & & & &  \\ 
 5000&  0.99&  -7.41$\pm $0.03&   1.54$\pm $0.02&   1.77$\pm $0.03&   1.62$\pm $0.03&   1.24$\pm $0.03&   0.87$\pm $0.02&  -0.01$\pm $0.00&  -0.22$\pm $0.00\\
& & & & & & & & &  \\ 
\hline
\hline
\end{tabular}
\end{table*}

\addtocounter{table}{-1}
\begin{table*}
\scriptsize
\caption{b): Broad band colours for the Y=0.27, Z=$10^{-3}$ model}
\begin{tabular}{ccccccccc}
\hline
\hline
& & & & & & & &   \\ 
age (Myr) &  V$_{\rm{tot}}$ & U-V & B-V & V-R & V-I & V-J & V-K & V-L \\
& & & & & & & &   \\ 
   30& -10.04&  -0.73$\pm $0.01&  -0.14$\pm $0.01&  -0.05$\pm $0.00&  -0.11$\pm $0.01&  -0.27$\pm $0.02&  -0.31$\pm $0.04&  -0.35$\pm $0.05\\
& & & & & & & &   \\ 
   50&  -9.81&  -0.58$\pm $0.02&  -0.10$\pm $0.01&  -0.03$\pm $0.01&  -0.07$\pm $0.01&  -0.17$\pm $0.03&  -0.17$\pm $0.04&  -0.19$\pm $0.04\\
& & & & & & & &   \\ 
   70&  -9.68&  -0.46$\pm $0.01&  -0.06$\pm $0.01&  -0.00$\pm $0.01&   0.00$\pm $0.01&   0.01$\pm $0.03&   0.12$\pm $0.05&   0.11$\pm $0.05\\
& & & & & & & &   \\ 
  100&  -9.52&  -0.35$\pm $0.01&  -0.03$\pm $0.01&   0.02$\pm $0.01&   0.05$\pm $0.01&   0.10$\pm $0.03&   0.24$\pm $0.04&   0.23$\pm $0.05\\
& & & & & & & &   \\ 
  150&  -9.32&  -0.22$\pm $0.01&   0.00$\pm $0.01&   0.04$\pm $0.01&   0.10$\pm $0.01&   0.22$\pm $0.02&   0.40$\pm $0.04&   0.41$\pm $0.04\\
& & & & & & & &   \\ 
  200&  -9.21&  -0.11$\pm $0.01&   0.04$\pm $0.00&   0.07$\pm $0.00&   0.15$\pm $0.01&   0.32$\pm $0.02&   0.53$\pm $0.02&   0.54$\pm $0.02\\
& & & & & & & &   \\ 
  300&  -8.97&   0.03$\pm $0.00&   0.10$\pm $0.00&   0.11$\pm $0.00&   0.23$\pm $0.01&   0.45$\pm $0.01&   0.70$\pm $0.02&   0.71$\pm $0.02\\
& & & & & & & &   \\ 
  400&  -8.77&   0.10$\pm $0.01&   0.15$\pm $0.00&   0.14$\pm $0.00&   0.29$\pm $0.01&   0.55$\pm $0.01&   0.82$\pm $0.02&   0.84$\pm $0.02\\
& & & & & & & &   \\ 
  500&  -8.64&   0.17$\pm $0.01&   0.21$\pm $0.00&   0.18$\pm $0.00&   0.38$\pm $0.01&   0.70$\pm $0.01&   1.01$\pm $0.02&   1.03$\pm $0.02\\
& & & & & & & &   \\ 
  600&  -8.66&   0.29$\pm $0.01&   0.29$\pm $0.01&   0.24$\pm $0.00&   0.50$\pm $0.01&   0.93$\pm $0.02&   1.35$\pm $0.03&   1.37$\pm $0.03\\
& & & & & & & &   \\ 
  800&  -8.47&   0.47$\pm $0.02&   0.40$\pm $0.01&   0.32$\pm $0.01&   0.65$\pm $0.02&   1.22$\pm $0.03&   1.76$\pm $0.04&   1.79$\pm $0.05\\
& & & & & & & &   \\ 
 1000&  -8.20&   0.49$\pm $0.02&   0.40$\pm $0.01&   0.32$\pm $0.01&   0.65$\pm $0.02&   1.22$\pm $0.03&   1.78$\pm $0.05&   1.81$\pm $0.05\\
& & & & & & & &   \\ 
 2000&  -7.79&   0.63$\pm $0.02&   0.49$\pm $0.01&   0.37$\pm $0.01&   0.74$\pm $0.01&   1.36$\pm $0.03&   1.96$\pm $0.03&   1.99$\pm $0.03\\
& & & & & & & &   \\ 
 3000&  -7.63&   0.69$\pm $0.03&   0.55$\pm $0.02&   0.40$\pm $0.01&   0.79$\pm $0.02&   1.44$\pm $0.04&   2.08$\pm $0.05&   2.11$\pm $0.05\\
& & & & & & & &   \\ 
 5000&  -7.41&   0.73$\pm $0.02&   0.59$\pm $0.01&   0.42$\pm $0.01&   0.83$\pm $0.02&   1.50$\pm $0.03&   2.15$\pm $0.04&   2.18$\pm $0.04\\
& & & & & & & &   \\ 
\hline
\hline
\end{tabular}
\end{table*}


\begin{table*}
\scriptsize
\caption{a): HST colours for the Y=0.23, Z=$10^{-3}$ model.}
\begin{tabular}{cccccccccc}
\hline
\hline
& & & & & & & & &  \\
age (Myr) & M$_{\rm{TO}}$ & V$_{\rm{tot}}$ & F152m-V & F170m-V & F253m-V & F278m-V & f307m-V & f480lp-V & f606w-V \\
& & & & & & & & &  \\
   30&  8.35&  -9.91$\pm $0.08&  -2.58$\pm $0.04&  -2.39$\pm $0.04&  -1.72$\pm $0.04&  -1.49$\pm $0.03&  -1.27$\pm $0.03&  -0.14$\pm $0.00&   0.02$\pm $0.00\\
& & & & & & & & &  \\
   50&  6.30&  -9.79$\pm $0.05&  -2.17$\pm $0.03&  -1.98$\pm $0.03&  -1.34$\pm $0.03&  -1.14$\pm $0.03&  -0.95$\pm $0.03&  -0.12$\pm $0.00&  -0.00$\pm $0.01\\
& & & & & & & & &  \\
   70&  5.32&  -9.65$\pm $0.05&  -1.90$\pm $0.02&  -1.73$\pm $0.02&  -1.12$\pm $0.01&  -0.94$\pm $0.01&  -0.77$\pm $0.01&  -0.12$\pm $0.00&  -0.01$\pm $0.01\\
& & & & & & & & &  \\
  100&  4.49&  -9.48$\pm $0.04&  -1.62$\pm $0.02&  -1.46$\pm $0.03&  -0.89$\pm $0.02&  -0.73$\pm $0.02&  -0.59$\pm $0.02&  -0.11$\pm $0.00&  -0.02$\pm $0.01\\
& & & & & & & & &  \\
  150&  3.72&  -9.30$\pm $0.04&  -1.28$\pm $0.03&  -1.14$\pm $0.02&  -0.61$\pm $0.02&  -0.49$\pm $0.02&  -0.38$\pm $0.02&  -0.10$\pm $0.01&  -0.04$\pm $0.01\\
& & & & & & & & &  \\
  200&  3.27&  -9.16$\pm $0.03&  -1.03$\pm $0.02&  -0.89$\pm $0.02&  -0.40$\pm $0.01&  -0.30$\pm $0.01&  -0.22$\pm $0.01&  -0.10$\pm $0.00&  -0.05$\pm $0.01\\
& & & & & & & & &  \\
  300&  2.75&  -8.92$\pm $0.03&  -0.65$\pm $0.03&  -0.54$\pm $0.03&  -0.10$\pm $0.02&  -0.04$\pm $0.02&  -0.00$\pm $0.02&  -0.09$\pm $0.00&  -0.06$\pm $0.00\\
& & & & & & & & &  \\
  400&  2.44&  -8.73$\pm $0.03&  -0.41$\pm $0.03&  -0.30$\pm $0.02&   0.09$\pm $0.02&   0.11$\pm $0.02&   0.11$\pm $0.02&  -0.08$\pm $0.00&  -0.08$\pm $0.01\\
& & & & & & & & &  \\
  500&  2.23&  -8.59$\pm $0.02&  -0.21$\pm $0.02&  -0.11$\pm $0.02&   0.23$\pm $0.02&   0.22$\pm $0.02&   0.19$\pm $0.01&  -0.08$\pm $0.01&  -0.09$\pm $0.00\\
& & & & & & & & &  \\
  600&  2.07&  -8.56$\pm $0.03&  -0.01$\pm $0.02&   0.11$\pm $0.02&   0.41$\pm $0.02&   0.36$\pm $0.02&   0.28$\pm $0.01&  -0.07$\pm $0.00&  -0.11$\pm $0.00\\
& & & & & & & & &  \\
  800&  1.86&  -8.38$\pm $0.02&   0.27$\pm $0.02&   0.41$\pm $0.02&   0.71$\pm $0.02&   0.62$\pm $0.01&   0.49$\pm $0.01&  -0.04$\pm $0.00&  -0.15$\pm $0.01\\
& & & & & & & & &  \\
 1000&  1.71&  -8.18$\pm $0.03&   0.41$\pm $0.02&   0.54$\pm $0.02&   0.82$\pm $0.03&   0.71$\pm $0.02&   0.56$\pm $0.02&  -0.04$\pm $0.00&  -0.16$\pm $0.00\\
& & & & & & & & &  \\
 2000&  1.35&  -7.69$\pm $0.04&   0.95$\pm $0.03&   1.02$\pm $0.04&   1.13$\pm $0.04&   0.93$\pm $0.03&   0.70$\pm $0.03&  -0.03$\pm $0.00&  -0.18$\pm $0.01\\
& & & & & & & & &  \\
 3000&  1.22&  -7.51$\pm $0.04&   1.28$\pm $0.02&   1.35$\pm $0.03&   1.31$\pm $0.03&   1.04$\pm $0.03&   0.76$\pm $0.02&  -0.02$\pm $0.00&  -0.20$\pm $0.01\\
& & & & & & & & &  \\
 5000&  1.07&  -7.21$\pm $0.04&   1.51$\pm $0.02&   1.71$\pm $0.03&   1.54$\pm $0.03&   1.16$\pm $0.03&   0.81$\pm $0.02&  -0.01$\pm $0.00&  -0.22$\pm $0.01\\
& & & & & & & & &  \\
\hline
\hline
\end{tabular}
\end{table*}

\addtocounter{table}{-1}
\begin{table*}
\scriptsize
\caption{b): Broad band colours for the Y=0.23, Z=$10^{-3}$ model.}
\begin{tabular}{ccccccccc}
\hline
\hline
& & & & & & & &   \\
age (Myr) &  V$_{\rm{tot}}$ & U-V & B-V & V-R & V-I & V-J & V-K & V-L \\
& & & & & & & &   \\
   30&  -9.91&  -0.75$\pm $0.02&  -0.14$\pm $0.01&  -0.05$\pm $0.01&  -0.12$\pm $0.01&  -0.29$\pm $0.01&  -0.33$\pm $0.02&  -0.38$\pm $0.02\\
& & & & & & & &   \\
   50&  -9.79&  -0.55$\pm $0.02&  -0.08$\pm $0.01&  -0.00$\pm $0.01&  -0.00$\pm $0.02&   0.02$\pm $0.06&   0.19$\pm $0.13&   0.18$\pm $0.13\\
& & & & & & & &   \\
   70&  -9.65&  -0.44$\pm $0.01&  -0.05$\pm $0.01&   0.02$\pm $0.01&   0.05$\pm $0.02&   0.13$\pm $0.06&   0.34$\pm $0.12&   0.34$\pm $0.12\\
& & & & & & & &   \\
  100&  -9.48&  -0.32$\pm $0.02&  -0.01$\pm $0.01&   0.04$\pm $0.01&   0.11$\pm $0.03&   0.26$\pm $0.07&   0.54$\pm $0.11&   0.55$\pm $0.12\\
& & & & & & & &   \\
  150&  -9.30&  -0.20$\pm $0.01&   0.03$\pm $0.01&   0.07$\pm $0.01&   0.16$\pm $0.02&   0.36$\pm $0.05&   0.65$\pm $0.08&   0.67$\pm $0.09\\
& & & & & & & &   \\
  200&  -9.16&  -0.10$\pm $0.01&   0.06$\pm $0.01&   0.09$\pm $0.01&   0.21$\pm $0.01&   0.45$\pm $0.02&   0.78$\pm $0.03&   0.80$\pm $0.03\\
& & & & & & & &   \\
  300&  -8.92&   0.04$\pm $0.01&   0.12$\pm $0.01&   0.13$\pm $0.01&   0.28$\pm $0.01&   0.59$\pm $0.03&   0.96$\pm $0.04&   0.98$\pm $0.04\\
& & & & & & & &   \\
  400&  -8.73&   0.12$\pm $0.01&   0.16$\pm $0.01&   0.15$\pm $0.01&   0.33$\pm $0.01&   0.66$\pm $0.03&   1.03$\pm $0.06&   1.05$\pm $0.06\\
& & & & & & & &   \\
  500&  -8.59&   0.17$\pm $0.01&   0.19$\pm $0.01&   0.17$\pm $0.01&   0.37$\pm $0.01&   0.70$\pm $0.03&   1.07$\pm $0.05&   1.09$\pm $0.05\\
& & & & & & & &   \\
  600&  -8.56&   0.24$\pm $0.01&   0.25$\pm $0.01&   0.22$\pm $0.01&   0.44$\pm $0.01&   0.83$\pm $0.02&   1.23$\pm $0.03&   1.25$\pm $0.04\\
& & & & & & & &   \\
  800&  -8.38&   0.42$\pm $0.01&   0.37$\pm $0.01&   0.30$\pm $0.01&   0.62$\pm $0.01&   1.15$\pm $0.02&   1.69$\pm $0.02&   1.72$\pm $0.02\\
& & & & & & & &   \\
 1000&  -8.18&   0.49$\pm $0.02&   0.40$\pm $0.01&   0.32$\pm $0.01&   0.65$\pm $0.01&   1.22$\pm $0.02&   1.78$\pm $0.04&   1.82$\pm $0.04\\
& & & & & & & &   \\
 2000&  -7.69&   0.59$\pm $0.02&   0.47$\pm $0.01&   0.35$\pm $0.01&   0.71$\pm $0.02&   1.32$\pm $0.04&   1.92$\pm $0.05&   1.96$\pm $0.05\\
& & & & & & & &   \\
 3000&  -7.51&   0.63$\pm $0.02&   0.52$\pm $0.01&   0.38$\pm $0.01&   0.75$\pm $0.02&   1.38$\pm $0.03&   1.99$\pm $0.05&   2.02$\pm $0.05\\
& & & & & & & &   \\
 5000&  -7.21&   0.69$\pm $0.02&   0.57$\pm $0.01&   0.41$\pm $0.01&   0.81$\pm $0.01&   1.45$\pm $0.03&   2.08$\pm $0.04&   2.11$\pm $0.04\\
& & & & & & & &   \\
\hline
\hline
\end{tabular}
\end{table*}


\begin{table*}
\scriptsize
\caption{Two colours calibration for the RF model.}
\begin{tabular}{ccc}
\hline
\hline
 & &  \\
age (Myr) & 15$-$31 & 18$-$28 \\
& &  \\
            8& -1.97& -1.26\\
& &  \\
           10& -1.90& -1.21\\
& &  \\
           15& -1.78& -1.12\\
& &  \\
           20& -1.70& -1.07\\
& &  \\
           25& -1.64& -1.04\\
& &  \\
           30& -1.55& -1.00\\
& &  \\
           50& -1.40& -0.91\\
& &  \\
           70& -1.25& -0.84\\
& &  \\
          100& -1.08& -0.76\\
& &  \\
          150& -0.86& -0.67\\
& &  \\
          200& -0.68& -0.60\\
& &  \\
          300& -0.21& -0.43\\
& &  \\
          400&  0.47& -0.25\\
& &  \\
          500&  1.59&  0.01\\
& &  \\
          600&  2.53&  0.19\\
& &  \\
          800&  4.74&  0.71\\
& &  \\
         1000&  6.54&  1.18\\
& &  \\
         1100&  7.21&  1.38\\
& &  \\
         1500&  8.84&  2.12\\
& &  \\
         1700&  9.45&  2.51\\
& &  \\
         2000& 10.11&  3.03\\
& &  \\
         3000& 10.71&  3.59\\
& &  \\
         4000& 11.71&  4.35\\
& &  \\
         5000& 12.12&  4.70\\
 & & \\
\hline
\hline
\end{tabular}
\end{table*}


\begin{table*}
\scriptsize
\caption{UV HST two colours calibration.}
\begin{minipage}[h]{10in}
\begin{tabular}{c|cc|cc|cc}
\hline
\hline
& & & & & & \\
 & \multicolumn{2}{|c|}{Z$=0.02$} & \multicolumn{2}{|c|}{Z$=0.006$} & \multicolumn{2}{|c}{Z$=0.001$} \\
& & & & & & \\
age (Myr) & C$_{1}$\footnote{C$_{1}$ is the F152M$-$F307M HST colour} & C$_{2}$\footnote{C$_{2}$ is the F170M$-$F278M HST colour} & C$_{1}^{~~a}$& $C_{2}^{~~b}$ & C$_{1}^{~~a}$ & C$_{2}^{~~b}$ \\
\hline
& & & & & & \\
             8 & -1.53 & -1.05 &       &       &       &      \\
& & & & & & \\
            10 & -1.47 & -1.00 &       &       &       &      \\
& & & & & & \\
            15 & -1.34 & -0.92 &       &       &       &      \\
& & & & & & \\
            20 & -1.26 & -0.88 &       &       &       &      \\
& & & & & & \\
            25 & -1.21 & -0.85 &       &       &       &      \\
& & & & & & \\
            30 & -1.15 & -0.81 & -1.20 & -0.84 & -1.29 & -0.89\\
& & & & & & \\
            50 & -1.02 & -0.74 & -1.08 & -0.77 & -1.21 & -0.84\\
& & & & & & \\
            70 & -0.92 & -0.68 & -1.00 & -0.72 & -1.12 & -0.78\\
& & & & & & \\
           100 & -0.79 & -0.61 & -0.90 & -0.66 & -1.01 & -0.72\\
& & & & & & \\
           150 & -0.64 & -0.53 & -0.78 & -0.59 & -0.90 & -0.65\\
& & & & & & \\
           200 & -0.54 & -0.48 & -0.67 & -0.52 & -0.79 & -0.58\\
& & & & & & \\
           300 & -0.29 & -0.34 & -0.50 & -0.43 & -0.60 & -0.47\\
& & & & & & \\
           400 & -0.01 & -0.21 & -0.37 & -0.35 & -0.47 & -0.38\\
& & & & & & \\
           500 &  0.26 & -0.03 & -0.24 & -0.28 & -0.37 & -0.30\\
& & & & & & \\
           600 &  0.45 &  0.14 & -0.10 & -0.20 & -0.27 & -0.24\\
& & & & & & \\
           800 &  0.63 &  0.37 &  0.18 & -0.03 & -0.20 & -0.21\\
& & & & & & \\
          1000 &  0.65 &  0.50 &  0.36 &  0.09 & -0.14 & -0.17\\
& & & & & & \\
          1100 &  0.65 &  0.54 &       &       &       &      \\
& & & & & & \\
          1500 &  0.70 &  0.67 &       &       &       &      \\
& & & & & & \\
          1700 &  0.69 &  0.70 &       &       &       &      \\
& & & & & & \\
          2000 &  0.67 &  0.72 &  0.67 &  0.48 &  0.27 &  0.11\\
& & & & & & \\
          3000 &  0.63 &  0.72 &  0.71 &  0.63 &  0.49 &  0.29\\
& & & & & & \\
          4000 &  0.61 &  0.72 &       &       &       &      \\
& & & & & & \\
          5000 &  0.58 &  0.69 &  0.71 &  0.75 &  0.67 &  0.54\\
& & & & & & \\
\hline
\hline
\end{tabular}
\end{minipage}
\end{table*}



\begin{thebibliography}{}
\bibitem[]{} {Alexander D. R., Brocato E., Cassisi S., Castellani V., Ciacio F., Degl'Innocenti S., 1997,  A\&A 317, 90}
\bibitem[]{} {Arimoto N., Yoshii Y., 1986, A\&A 164, 260}
\bibitem[]{} {Arimoto N., Yoshii Y., 1987, A\&A 173, 23}
\bibitem[]{} {Barbaro G., Olivi F.M., 1986, in Spectral Evolution of
Galaxies ed. Chiosi C., Renzini A., Reidel, p. 283} 
\bibitem[]{}{Barbero J., Brocato E., Cassatella A., Castellani V., Geyer E. H.,1990, ApJ 351, 98}
\bibitem[]{}{Bencivenni D., Brocato E., Buonanno R., Castellani V.,
1991 AJ, 102, 137 }
\bibitem[]{}{Bl\"{o}cker T., Sch\"{o}nberner G. 1991, A\&A 244, L43}
\bibitem[]{}{Bressan A., Chiosi C., Fagotto F.
1994 ApJS 94, 64 (B94)}
\bibitem[]{}{Brocato E., Buonanno R., Castellani V., Walker A. R., 1989, ApJSS 71, 25} 
\bibitem[]{}{Brocato E., Matteucci F., Mazzitelli I., Tornamb\`e A., 1990, ApJ 349, 458}
\bibitem[]{}{Brocato E., Castellani V., 1993, ApJ 410, 99}
\bibitem[]{}{Brocato E., Castellani V., Di Giorgio A., 1993, AJ 105, 2192}
\bibitem[]{}{Brocato E., Castellani V. Piersimoni A., 1994 A\&A 290, 59}
\bibitem[]{}{Brocato E., Castellani V. Piersimoni A., 1997, ApJ 491, 789}
\bibitem[]{}{Bruzual G.A., Charlot S., 1993, ApJ 405, 538}
\bibitem[]{}{Cassisi S., Castellani V., Straniero O., 1994, A\&A  282, 753 }
\bibitem[]{}{Caputo F., Chieffi A., Castellani V., Collados M., 
Martinez Roger C., 1990, AJ 99, 261}
\bibitem[]{}{Cassatella A., Barbero J., Brocato E., Castellani V., Geyer E.H. 1996 A\&A 306, 125}
\bibitem[]{}{Castellani V., Chieffi A. Straniero O., 1992, ApJSS 78, 517 }
\bibitem[]{}{Charlot S., Worthey G., Bressan A., 1996, ApJ 457, 625}
\bibitem[]{}{Chiosi C., Vallenari A., Bressan A., 1997, A\&AS 121, 301}
\bibitem[]{}{Dorman B., Rood R.T., O'Connell R.W., 1993, ApJ 419, 596}
\bibitem[]{}{Dorman B., O'Connell R.W., Rood R.T., 1995, ApJ 442, 105}
\bibitem[]{}{Ferraro F.R., Fusi Pecci F., Testa V., Greggio L., Corsi C.E., Buonanno R., Terndrup D.M., Zinneker H., 1995, MNRAS 272, 391}
\bibitem[]{}{Gilmozzi R., Kinney E.K., Ewald S.P., Panagia N., 
Romaniello M., 1994, ApJ 435, L43}
\bibitem[]{}{Holtzman J. A., Faber S. M., Shaya E. J., et al., 1992, AJ, 103, 691}
\bibitem[]{}{Kurucz R.L. 1979a, ApJSS, 49, 1 (K79)}
\bibitem[]{}{Kurucz R.L. 1979b: in "Problems of Calibration of Multicolour Photometric
Systems", ed. A.G. Davis Philip, Dudley Obs. Rep. No. 14, p. 363}
\bibitem[]{}{Kurucz R.L. 1992, in IAU Symposium 149 "The Stellar Populations
of Galaxies", ed. B. Barbuy \& A. Renzini (Dordrecht: Kluwer), 225}
\bibitem[]{}{Larson R.B. 1974, MNRAS 166, 585}
\bibitem[]{}{Leitherer C., Heckman T.M., 1995, ApJSS 96, 9}
\bibitem[]{}{Leitherer C., Alloin D., Alvensleben U., et al. 1996, PASP 108, 996}
\bibitem[]{}{Maeder A., Meynet G., 1989, A\&A 210, 155}
\bibitem[]{}{Maeder A., Meynet G., 1991, A\&AS 89, 451}
\bibitem[]{}{Renzini A., Buzzoni A. 1986, in Spectral Evolution of Galaxies
eds. Chiosi C., Renzini A., Reidel, p. 195}
\bibitem[]{}{Renzini A. 1992, in The Stellar Population of Galaxies, ed. B. 
Barbuy and A. Renzini, Dordrecht: Kluwer, p. 325}
\bibitem[]{} {Rocca-Volmerange B., Guiderdoni B., 1987, A\&A 175, 15}
\bibitem[]{} {Salpeter E. E., 1955, ApJ 121, 161}
\bibitem[]{}{Searle L., Wilkinson A., Bagnuolo W.G., 1980, ApJ 239, 803}
\bibitem[]{}{Straniero O., Chieffi A., 1991, ApJSS 76, 525}
\bibitem[]{}{Stothers R. B., Chin C., 1992, ApJ 390, 136}
\bibitem[]{}{Schweizer F., Miller B.W., Whitmore B. C., Fall S.M., 1996,
AJ 112, 1839}
\bibitem[]{}{Tinsley B.M., 1980, ApJ 243, 41}
\bibitem[]{}{Vazdekis A., Casuso E., Peletier R., Beckman J.E., 1996, 
ApJSS 106,307}
\bibitem[]{}{Yi S., Demarque P., Oemler A. Jr., 1995, PASP 107, 273}
\bibitem[]{}{Worthey G., 1994, ApJSS 95, 107}
\end{thebibliography}
\end{document}